\def\be{\begin{equation}}
\def\ee{\end{equation}}
\def\ba{\begin{array}}
\def\ea{\end{array}}

\documentclass[prl,showpacs,twocolumn,amsmath]{revtex4}
\usepackage{amsfonts}
\usepackage{mathrsfs}
\usepackage{mathdots}
\usepackage{amssymb}
\usepackage{pifont}
\usepackage{epsfig,subfigure,dsfont,amsthm,amsbsy,mathrsfs,amscd}

\usepackage{rotating}
\UseRawInputEncoding
\usepackage{graphicx}
\usepackage{dcolumn}
\usepackage{bm}
\usepackage{amsmath}
\usepackage{multirow}
\usepackage{array}
\usepackage{epstopdf}
\usepackage{diagbox}
\usepackage{dcolumn}

\def\qed{\leavevmode\unskip\penalty9999 \hbox{}\nobreak\hfill
     \quad\hbox{\leavevmode  \hbox to.77778em{%
               \hfil\vrule   \vbox to.675em%
               {\hrule width.6em\vfil\hrule}\vrule\hfil}}
     \par\vskip3pt}

\newtheorem{lemma}{Lemma}

\begin{document}

\title{Tripartite quantum steering in Schwarzschild spacetime}
\author{Guang-Wei Mi$^{1}$}
\author{Xiaofen Huang $^{1}$}
\author{Tinggui Zhang$^{1,*}$}
\affiliation{$^{1}$School of Mathematics and Statistics, Hainan Normal University, Haikou, 571158, China\\
{$^{*}$ Email address: tinggui333@163.com}}


\begin{abstract}
We investigate the effects of Hawking radiation on quantum steering and steering asymmetry in a tripartite system embedded in Schwarzschild spacetime. All tripartite steering types were classified, comprising three $1\rightarrow2$ and three $2\rightarrow1$ steering cases. Through a systematic analysis of all physically relevant scenarios (including accessible and inaccessible modes), we classify three canonical scenarios with one, two and three physically accessible modes. In the scenario of three physically accessible modes, Hawking radiation disrupts quantum steering, with the maximum steering asymmetry during the two-way steering to one-way steering transition precisely demarcating the phase boundary between these regimes.  For two physically accessible modes, Hawking radiation exhibits dual behavior: enhancing the steering from Alice and Bob to anti-Charlie under certain parameters while suppressing it under others, while net strengthening other steering types.  When considering one physically accessible mode, the Hawking effect of the black hole significantly enhances quantum steering. These findings provide new insights into quantum correlations in curved spacetime and establish observable signatures of Hawking effects in quantum steering phenomena.
\end{abstract}

\maketitle


\section{Introduction}
Quantum steering, first conceptualized by Schr\"{o}dinger in 1935 as an extension of the Einstein-Podolsky-Rosen (EPR) paradox, characterizes the capacity of one observer to nonlocally manipulate the quantum states of a distant system via local measurements~\cite{Einstein.1935,Schrodinger.1935,Schrodinger.1936}. In 2007, since Wiseman \emph{et al.} formalized quantum steering through the local hidden state (LHS) model and established its operational hierarchy as a quantum correlation intermediate between entanglement and Bell nonlocality~\cite{Wiseman.2007}, the field has witnessed renewed research interest~\cite{Walborn.2011,Navascues.2012,Skrzypczyk.2014,Deng.2017,Nery.2018,Nery.2020,Designolle.2021}. Within this framework, a variety of inequalities have been proposed for detecting quantum steering in quantum states~\cite{Chen.2014,Kogias.2015,Ren.2018,Maity.2018}. Furthermore, experimental evidence has confirmed the asymmetry of quantum steering in quantum systems~\cite{Saunders.2010,Wollmann.2016,Xiao.2017}. Due to quantum steering asymmetry, bipartite systems exhibit three distinct asymmetric configurations: two-way, one-way, and no-way steering. The distinct directional properties of EPR steering have been extensively studied and implemented in fundamental quantum information tasks and asymmetric quantum communication schemes~\cite{Branciard.2012,Handchen.2012,Armstrong.2015,Tischler.2018,Fan.2022}.

In the early 20th century, Albert Einstein established the theory of general relativity, reconceptualizing gravity as a geometric consequence of spacetime curvature dynamically coupled to mass-energy distributions. In 1916, Karl Schwarzschild~\cite{Schwarzschild.1916} derived the first exact solution to Einstein’s field equations, defining the Schwarzschild radius as a critical boundary below which static spherically symmetric masses exhibit inescapable gravitational pull for light, thereby providing the first rigorous mathematical confirmation of black hole existence. In 1974, Stephen Hawking proposed the Hawking radiation theory~\cite{Hawking.1974}, establishing a profound connection between black holes and quantum mechanics, which subsequently led to the formulation of the black hole information paradox~\cite{Hawking.1975,Hawking.1976,Bombelli.1986}. Entering the 21st century, humanity achieved a remarkable theoretical-to-empirical leap, with the 2019 imaging of M878*~\cite{EHT.L1,EHT.L2,EHT.L3,EHT.L4,EHT.L5,EHT.L6} and the 2022 release of Sgr A*'s~\cite{EHT.L17} observational data providing further validation of general relativity.

In recent years, driven by the pursuit of reconciling quantum mechanics with general relativity, the investigation of quantum information in non-inertial reference frames and curved spacetimes has emerged as a rapidly evolving research frontier~\cite{Schuller.2005,Pan.2008,Wang.2009,Wang.2010,Esfahani.2011,Bhattacharya.2022,Wu.2022,Li.2022,Zhang.2023,Mi.2025,Dong.2018,Dong.2019,Dong.2025}. In~\cite{D.Das.2019}, Debarshi Das\emph{et al.} propose a criterion to detect whether a given two-qubit state is EPR steerable by constructing another two-qubit state from it, such that the original state is EPR steerable if the newly constructed state is entangled. In~\cite{EPJC.2022}, Wu \emph{et al.} investigated the effects of Hawking radiation on quantum steering and its monogamy constraints for fermionic modes in Schwarzschild spacetime. In~\cite{Fei.2022}, Chen \emph{et al.} present effective criteria for (genuine) tripartite steerability of three-qubit states by linking it to the (genuine) tripartite entanglement of corresponding quantum states. In~\cite{Wu.2025}, the authors study tripartite quantum steering dynamics in a three-Unruh-DeWitt-detector system under the GHZ and W states.

In the paper, we explore the influence of Hawking effects on quantum steering in a tripartite system within Schwarzschild spacetime. Assume that Alice, Bob and Charlie initially share a GHZ state at an asymptotically flat region. Then, let Alice remains in the asymptotically flat region, while Bob and Charlie are positioned as static observers in the vicinity of the black hole's event horizon. Next, we conduct a comprehensive analysis of all feasible configurations, encompassing both physically accessible and inaccessible modes, and explicitly delineate three distinct scenarios: (i) three physically accessible modes, (ii) two physically accessible modes, and (iii) one physically accessible mode.
In the first scenario, the Hawking effect disrupts quantum steering. The transition from two-way to one-way steering marked by maximal steering asymmetry precisely defines the phase boundary between these regimes in black hole spacetime. In the second scenario, the Hawking effect exhibits a dual influence on quantum steering: it enhances steering $S^{AB\rightarrow c}_{H}$ while suppressing it under specific parameter regimes. For other steering types, the Hawking effect yields a net enhancement. In the third scenario, the Hawking effect enhances quantum steering. Interestingly, across all three considered scenarios, the $1\rightarrow2$ steering strength consistently surpasses that of the $2\rightarrow1$ steering.

The rest of this paper is organized as follows. In Sec.\uppercase\expandafter{\romannumeral2}, we provided a concise overview of the Dirac field quantization in Schwarzschild spacetime. In Sec.\uppercase\expandafter{\romannumeral3}, we quantified all types of tripartite quantum steering for X-type states through entanglement.
In Sec. \uppercase\expandafter{\romannumeral4}, we investigate the impact of Hawking radiation on tripartite quantum steering in Schwarzschild spacetime.
We conclude in Sec.\uppercase\expandafter{\romannumeral5}.

\section{QUANTIZATION OF DIRAC FIELD IN SCHWARZSCHILD BLACK HOLE}

The Dirac equation within a generic background spacetime is capable of being formulated as~\cite{Dirac.1957},
\begin{eqnarray}
\begin{aligned}
\left[\gamma^{a}e^{\mu}_{a}(\partial_{\mu}+\Gamma_{\mu})+\mu\right]\Phi=0,
\end{aligned}
\end{eqnarray}
where $\mu$ is the mass of the Dirac field, $\gamma^{a}$ is the Dirac matrix, $e^{\mu}_{a}$ is the inverse of the tetrad $e^{a}_{\mu}$ and $\Gamma_{\mu}=\frac{1}{8}[\gamma^{a},\gamma^{b}]e^{\nu}_{a}e_{b\nu}$ is the spin connection.
The metric characterizing Schwarzschild spacetime can be written as~\cite{metric.2010}
\begin{eqnarray}
\begin{aligned}
ds^{2}=&-\left(1-\frac{2M}{r}\right)dt^{2}+\left(1-\frac{2M}{r}\right)^{-1}dr^{2}\\[1mm]
&+r^{2}(d\theta^{2}+\sin^{2}\theta d\varphi^{2}),
\end{aligned}
\end{eqnarray}
where $M$ is the mass of the black hole. 
Specifically, the massless Dirac equation in Schwarzschild spacetime takes the form as~\cite{Xu.2014}
\begin{eqnarray}
\small
\label{SchDirac}
\begin{aligned}
&-\frac{\gamma_{0}}{\sqrt{1-\frac{2M}{r}}}\frac{\partial\Phi}{\partial t}+\gamma_{1}\sqrt{1-\frac{2M}{r}}\left[\frac{\partial}{\partial r}+\frac{1}{r}+\frac{M}{2r(r-2M)}\right]\Phi\\
&+\frac{\gamma_{2}}{r}\left(\frac{\partial}{\partial\theta}
+\frac{\cot\theta}{2}\right)\Phi
+\frac{\gamma_{3}}{r\sin\theta}\frac{\partial\Phi}{\partial\varphi}=0.
\end{aligned}
\end{eqnarray}

Eq.(\ref{SchDirac}) explicitly constructs positive frequency outgoing modes that permeate both the interior and exterior regions of the event horizon~\cite{Wang.2010,Jing.2004,metric.2010},
\begin{eqnarray}
\label{kout}
\Phi^{+}_{k,out}=\phi(r)e^{-i\omega\mu},
\end{eqnarray}
\begin{eqnarray}
\label{kin}
\Phi^{+}_{k,in}=\phi(r)e^{i\omega\mu},
\end{eqnarray}
where $k$ is the wave vector, $\phi(r)$ represents the four-component Dirac spinor, $\omega$ indicates a monochromatic frequency and $\mu=t-r_{*}$ with the tortoise
coordinate $r_{*}=r+2M\ln\frac{r-2M}{2M}$.
Moreover, the Dirac field $\Phi$ admits mode expansion through Eqs.(\ref{kout}) and (\ref{kin}) as
\begin{eqnarray}
\begin{aligned}
\label{Sphi}
\Phi&=\int dk\left[\hat{a}^{in}_{k}\Phi^{+}_{k,in}+\hat{b}^{in\dag}_{-k}\Phi^{-}_{-k,in}\right.\\
&+\left.\hat{a}^{out}_{k}\Phi^{+}_{k,out}+\hat{b}^{out\dag}_{-k}\Phi^{-}_{-k,out}\right],
\end{aligned}
\end{eqnarray}
where $\hat{a}^{in}_{k}$ corresponds to the fermionic annihilation operator in the interior of the event horizon, whereas $\hat{b}^{out\dag}_{-k}$ represents the antifermionic creation operator in the exterior region. Generally, the modes $\Phi^{\pm}_{k,out}$ and $\Phi^{\pm}_{k,in}$ are conventionally designated as Schwarzschild modes.

Following the framework proposed by Damour and Ruffini~\cite{Damoar.1976}, we construct a complete orthonormal basis for the positive-energy Kruskal modes, thereby explicitly linking Eqs.(\ref{kout}) and (\ref{kin})
\begin{eqnarray}
\label{Kout}
\Psi^{+}_{k,out}=e^{-2\pi M\omega}\Phi^{-}_{-k,in}+e^{2\pi M\omega}\Phi^{+}_{k,out},
\end{eqnarray}
\begin{eqnarray}
\label{Kin}
\Psi^{+}_{k,in}=e^{-2\pi M\omega}\Phi^{-}_{-k,out}+e^{2\pi M\omega}\Phi^{+}_{k,in}.
\end{eqnarray}
Hence, the Dirac field $\Phi$ admits expansion in terms of Kruskal modes
\begin{eqnarray}
\begin{aligned}
\label{Kphi}
\Phi&=\int dk\left[2\cosh(4\pi M\omega)\right]^{-\frac{1}{2}}\left[\hat{c}^{in}_{k}\Psi^{+}_{k,in}\right.\\
&+\left.\hat{d}^{in\dag}_{-k}\Psi^{-}_{-k,in}+\hat{c}^{out}_{k}\Psi^{+}_{k,out}+\hat{d}^{out\dag}_{-k}\Psi^{-}_{-k,out}\right],
\end{aligned}
\end{eqnarray}
where $\hat{c}^{\theta}_{k}$ and $\hat{d}^{\theta}_{k}$ ($\theta=(in,out)$) denote the fermionic annihilation operators along with the antifermionic creation operators, both acting on the Kruskal vacuum.

Consequently, Eqs.(\ref{Sphi}) and (\ref{Kphi}) reveal that the Dirac field undergoes quantization through the Schwarzschild and Kruskal operators, respectively. This process induces a Bogoliubov transformation~\cite{Barnett.1997} between the two operator sets, with its explicit form detailed as follows
\begin{eqnarray}
\begin{aligned}
\hat{c}^{out}_{k}=\frac{1}{\sqrt{e^{-8\pi M\omega}+1}}\hat{a}^{out}_{k}-\frac{1}{\sqrt{e^{8\pi M\omega}+1}}\hat{b}^{in\dag}_{-k},
\end{aligned}
\end{eqnarray}
\begin{eqnarray}
\begin{aligned}
\hat{c}^{out\dag}_{k}=\frac{1}{\sqrt{e^{-8\pi M\omega}+1}}\hat{a}^{out\dag}_{k}-\frac{1}{\sqrt{e^{8\pi M\omega}+1}}\hat{b}^{in}_{-k}.
\end{aligned}
\end{eqnarray}
Through the application of Bogoliubov transformations, the Kruskal vacuum and excited states in Schwarzschild spacetime can be expressed as~\cite{Wang2010,Phy2024},
\begin{eqnarray}
\begin{aligned}
\label{eqK}
&|0\rangle_{K}=\frac{1}{\sqrt{e^{-\frac{\omega}{T}}+1}}|0\rangle_{out}|0\rangle_{in}+\frac{1}{\sqrt{e^{\frac{\omega}{T}}+1}}|1\rangle_{out}|1\rangle_{in},\\
&|1\rangle_{K}=|1\rangle_{out}|0\rangle_{in},
\end{aligned}
\end{eqnarray}
where $T=\frac{1}{8\pi M}$ is the Hawking temperature, $|n\rangle_{out}$ and $|n\rangle_{in}$ correspond to the fermionic modes outside the event horizon and the antifermionic modes inside the event horizon, respectively.

\section{Quantification of the Tripartite Quantum Steering}
In this section, we conduct a systematic analysis of tripartite quantum steering, quantifying all possible steering scenarios, including three types of $1\rightarrow2$ steering and three types of $2\rightarrow1$ steering. Consider the following X-type state, whose density matrix takes the form of
\begin{eqnarray}
\label{rhox}
\rho_{X}=
\left (
\begin{array}{cccccccc}
\rho_{11}   &0               &0                  &0                 &0                &0                     &0                       &\rho_{18}\\
0           &\rho_{22}       &0                  &0                 &0                &0                     &\rho_{27}               &0        \\
0           &0               &\rho_{33}          &0                 &0                &\rho_{36}             &0                       &0        \\
0           &0               &0                  &\rho_{44}         &\rho_{45}        &0                     &0                       &0        \\
0           &0               &0                  &\rho_{54}         &\rho_{55}        &0                     &0                       &0        \\
0           &0               &\rho_{63}          &0                 &0                &\rho_{66}             &0                       &0        \\
0           &\rho_{72}       &0                  &0                 &0                &0                     &\rho_{77}               &0        \\
\rho_{81}   &0               &0                  &0                 &0                &0                     &0                       &\rho_{88}
\end{array}
\right ),
\end{eqnarray}
where $\rho_{11}+\rho_{22}+\rho_{33}+\rho_{44}+\rho_{55}+\rho_{66}+\rho_{77}+\rho_{88}=1$, $\rho_{ii}\rho_{jj}\geq|\rho_{ij}|^{2}$$(i\neq j$ and $i+j=9)$, and $\rho_{ij}=\rho_{ji}$.
If $\rho_{44}\rho_{55}<|\rho_{18}|^{2}$ or $\rho_{33}\rho_{66}<|\rho_{27}|^{2}$ or $\rho_{22}\rho_{77}<|\rho_{36}|^{2}$ or $\rho_{11}\rho_{88}<|\rho_{45}|^{2}$, the X-type state $\rho_{X}$ is entangled.

Prior to quantifying tripartite quantum steering, we introduce the following lemmas.
\begin{lemma}(Observation 3 of Ref.~\cite{D.Das.2019})
\label{lemma111}
For any qudit-qubit state $\rho_{AB}$ shared between Alice and Bob, define another new qudit-qubit state $\tau_{AB}$ given by
\begin{eqnarray}
\begin{aligned}
\tau^{1}_{AB}=\mu_{1}\rho_{AB}+(1-\mu_{1})\tilde{\rho^{1}}_{AB},
\end{aligned}
\end{eqnarray}
where $\tilde{\rho^{1}}_{AB}=\rho_{A}\otimes \frac{\mathbb{I}_{2}}{2}$ with $\rho_{A}=Tr_{B}[\rho_{AB}]=Tr_{B}[\tau^{1}_{AB}]$ being the reduced state (qudit) at Alice’s side and $\mathbb{I}_{2}$ is the $2\times2$ identity matrix.
If $\tau^{1}_{AB}$ is entangled, then $\rho_{AB}$ is EPR steerable from Bob to Alice for $0\leq\mu_{1}\leq\frac{1}{\sqrt{3}}$.
\end{lemma}

\begin{lemma}(Theorem 2 of Ref.~\cite{Fei.2022})
\label{lemma222}
Let $\rho_{ABC}$ be a three-qubit state and
\begin{eqnarray}
\begin{aligned}
\tau^{2}_{ABC}=\mu_{2}\rho_{ABC}+(1-\mu_{2})\frac{\mathbb{I}_{4}}{4}\otimes\rho_{C},
\end{aligned}
\end{eqnarray}
where $\rho_{C}=Tr_{AB}[\rho_{ABC}]$ and $\mathbb{I}_{4}$ is the $4\times4$ identity matrix.
If $\tau^{2}_{ABC}$ is tripartite entangled, then $\rho_{ABC}$ is tripartite steerable from Alice and Bob to Charlie for $0\leq\mu_{2}\leq\frac{1}{3}$.
\end{lemma}
Note that in Lemma \ref{lemma111}, one can detect the EPR steering from Bob to Alice for the qudit-qubit state shared between Alice and Bob, but it does not enable the detection of the EPR steering from Alice to Bob for the same qudit-qubit state shared between them.

For a tripartite state ${\rho}_{ABC}$ shared by Alice, Bob, and Charlie, it can be decomposed into the following three types of bipartite partitions as $AB|C$, $BC|A$, $CA|B$.
Then, according to Lemma \ref{lemma111}, one can detect the steering from Charlie to Alice and Bob when the following state
\begin{eqnarray}
\begin{aligned}
\label{ABC}
\tau^{1}_{AB|C}=\frac{\rho_{ABC}}{\sqrt{3}}+\frac{3-\sqrt{3}}{3}\left({\rho}_{AB}\otimes \frac{\mathbb{I}_{2}}{2}\right)
\end{aligned}
\end{eqnarray}
is entangled, where $\rho_{AB}=Tr_{C}[\rho_{ABC}]$ and $\mu_{1}=\frac{1}{\sqrt{3}}$ in Lemma \ref{lemma111}.
Additionally, in accordance with Lemma \ref{lemma222}, the steering from Alice and Bob to Charlie can be discerned when considering the following state
\begin{eqnarray}
\begin{aligned}
\label{ABC2}
\tau^{2}_{AB|C}=\frac{\rho_{ABC}}{3}+\frac{2}{3}\left(\frac{\mathbb{I}_{4}}{4}\otimes\rho_{C}\right)
\end{aligned}
\end{eqnarray}
is entangled, where $\rho_{C}=Tr_{AB}[\rho_{ABC}]$ and $\mu_{2}=\frac{1}{3}$ in Lemma \ref{lemma222}.

Similarly, the steering from A to BC, BC to A, B to CA and CA to B can be witnessed when the steered density matrix $\tau^{1}_{BC|A}$, $\tau^{2}_{BC|A}$, $\tau^{1}_{CA|B}$ and $\tau^{2}_{CA|B}$ defined as
\begin{eqnarray}
\begin{aligned}
\label{BCA}
\tau^{1}_{BC|A}=\frac{\rho_{BCA}}{\sqrt{3}}+\frac{3-\sqrt{3}}{3}\left({\rho}_{BC}\otimes \frac{\mathbb{I}_{2}}{2}\right),
\end{aligned}
\end{eqnarray}
\begin{eqnarray}
\begin{aligned}
\label{BCA2}
\tau^{2}_{BC|A}=\frac{\rho_{BCA}}{3}+\frac{2}{3}\left(\frac{\mathbb{I}_{4}}{4}\otimes\rho_{A}\right),
\end{aligned}
\end{eqnarray}
\begin{eqnarray}
\begin{aligned}
\label{CAB}
\tau^{1}_{CA|B}=\frac{\rho_{CAB}}{\sqrt{3}}+\frac{3-\sqrt{3}}{3}\left({\rho}_{CA}\otimes \frac{\mathbb{I}_{2}}{2}\right),
\end{aligned}
\end{eqnarray}
\begin{eqnarray}
\begin{aligned}
\label{CAB2}
\tau^{2}_{CA|B}=\frac{\rho_{CAB}}{3}+\frac{2}{3}\left(\frac{\mathbb{I}_{4}}{4}\otimes\rho_{B}\right)
\end{aligned}
\end{eqnarray}
are entangled, respectively. Here, $\rho_{BC}=Tr_{A}[\rho_{BCA}]$, $\rho_{A}=Tr_{BC}[\rho_{BCA}]$, $\rho_{CA}=Tr_{B}[\rho_{CAB}]$ and $\rho_{B}=Tr_{CA}[\rho_{CAB}]$.

Without loss of generality, let $\rho_{ABC}=\rho_{X}$. Therefore, we derive that the steered density matrix $\tau^{1}_{AB|C}$ of $\rho_{ABC}$ can be explicitly expressed as
\begin{widetext}
\small
\begin{eqnarray}
\label{tauABC}
\tau^{1}_{AB|C}=
\left (
\begin{array}{cccccccc}
\frac{\sqrt{3}}{3}\rho_{11}+m_{1}  &0               &0                  &0                 &0                &0                     &0            &\frac{\sqrt{3}}{3}\rho_{18}\\
0            &\frac{\sqrt{3}}{3}\rho_{22}+m_{1}     &0                  &0                 &0                &0                     &\frac{\sqrt{3}}{3}\rho_{27}           &0\\
0            &0               &\frac{\sqrt{3}}{3}\rho_{33}+m_{2}          &0                 &0                &\frac{\sqrt{3}}{3}\rho_{36}             &0                 &0\\
0            &0               &0            &\frac{\sqrt{3}}{3}\rho_{44}+m_{2}       &\frac{\sqrt{3}}{3}\rho_{45}          &0                     &0                       &0\\
0            &0               &0            &\frac{\sqrt{3}}{3}\rho_{54}         &\frac{\sqrt{3}}{3}\rho_{55}+m_{3}        &0                     &0                       &0\\
0            &0               &\frac{\sqrt{3}}{3}\rho_{63}    &0                 &0                &\frac{\sqrt{3}}{3}\rho_{66}+m_{3}             &0                       &0\\
0            &\frac{\sqrt{3}}{3}\rho_{72}       &0            &0                 &0                &0                  &\frac{\sqrt{3}}{3}\rho_{77}+m_{4}                  &0\\
\frac{\sqrt{3}}{3}\rho_{81}    &0               &0            &0                 &0                &0                  &0                     &\frac{\sqrt{3}}{3}\rho_{88}+m_{4}
\end{array}
\right ),
\end{eqnarray}
\end{widetext}
where $m_{1}=\frac{3-\sqrt{3}}{6}(\rho_{11}+\rho_{22})$, $m_{2}=\frac{3-\sqrt{3}}{6}(\rho_{33}+\rho_{44})$, $m_{3}=\frac{3-\sqrt{3}}{6}(\rho_{55}+\rho_{66})$, and $m_{4}=\frac{3-\sqrt{3}}{6}(\rho_{77}+\rho_{88})$.
The state $\tau^{1}_{AB|C}$ is entangled provided that it satisfies the inequalities
\begin{eqnarray}
\begin{aligned}
\label{Pa}
|\rho_{18}|^{2}>P^{1}_{a,-} \quad \quad \quad or \quad \quad \quad |\rho_{27}|^{2}>P^{1}_{a,+},
\end{aligned}
\end{eqnarray}
or
\begin{eqnarray}
\begin{aligned}
\label{Pc}
|\rho_{36}|^{2}>P^{1}_{b,-} \quad \quad \quad or \quad \quad \quad |\rho_{45}|^{2}>P^{1}_{b,+},
\end{aligned}
\end{eqnarray}
where
\begin{eqnarray*}
\begin{aligned}
P^{1}_{a,\pm}=&\frac{2\pm\sqrt{3}}{2}\rho_{33}\rho_{66}+\frac{2\mp\sqrt{3}}{2}\rho_{44}\rho_{55}\\
&+\frac{1}{2}(\rho_{33}\rho_{55}+\rho_{44}\rho_{66}),\\
\end{aligned}
\end{eqnarray*}
\begin{eqnarray}
\begin{aligned}
\label{Pa-d}
P^{1}_{b,\pm}=&\frac{2\pm\sqrt{3}}{2}\rho_{11}\rho_{88}+\frac{2\mp\sqrt{3}}{2}\rho_{22}\rho_{77}\\
&+\frac{1}{2}(\rho_{11}\rho_{77}+\rho_{22}\rho_{88}).
\end{aligned}
\end{eqnarray}
Thus, the steering from Charlie to Alice and Bob is observed. Next, we deduce that the steered density matrix $\tau^{2}_{AB|C}$ pertaining to $\rho_{ABC}$ can be precisely formulated as
\begin{widetext}
\begin{eqnarray}
\label{tau2ABC}
\tau^{2}_{AB|C}=
\left (
\begin{array}{cccccccc}
\frac{1}{3}\rho_{11}+n_{1}  &0               &0                  &0                 &0                &0                     &0            &\frac{1}{3}\rho_{18}\\
0            &\frac{1}{3}\rho_{22}+n_{2}     &0                  &0                 &0                &0                     &\frac{1}{3}\rho_{27}           &0\\
0            &0               &\frac{1}{3}\rho_{33}+n_{1}          &0                 &0                &\frac{1}{3}\rho_{36}             &0                 &0\\
0            &0               &0            &\frac{1}{3}\rho_{44}+n_{2}       &\frac{1}{3}\rho_{45}          &0                     &0                       &0\\
0            &0               &0            &\frac{1}{3}\rho_{54}         &\frac{1}{3}\rho_{55}+n_{1}        &0                     &0                       &0\\
0            &0               &\frac{1}{3}\rho_{63}    &0                 &0                &\frac{1}{3}\rho_{66}+n_{2}             &0                       &0\\
0            &\frac{1}{3}\rho_{72}       &0            &0                 &0                &0                  &\frac{1}{3}\rho_{77}+n_{1}                  &0\\
\frac{1}{3}\rho_{81}    &0               &0            &0                 &0                &0                  &0                     &\frac{1}{3}\rho_{88}+n_{2}
\end{array}
\right ),
\end{eqnarray}
\end{widetext}
where $n_{1}=\frac{1}{6}(\rho_{11}+\rho_{33}+\rho_{55}+\rho_{77})$ and $n_{2}=\frac{1}{6}(\rho_{22}+\rho_{44}+\rho_{66}+\rho_{88})$.
The state $\tau^{2}_{AB|C}$ is entangled whenever it fulfills the inequalities
\begin{eqnarray}
\begin{aligned}
\label{Pa2}
|\rho_{18}|^{2}>P^{2}_{a,-} \quad \quad \quad or \quad \quad \quad |\rho_{27}|^{2}>P^{2}_{a,+},
\end{aligned}
\end{eqnarray}
or
\begin{eqnarray}
\begin{aligned}
\label{Pc2}
|\rho_{36}|^{2}>P^{2}_{b,-} \quad \quad \quad or \quad \quad \quad |\rho_{45}|^{2}>P^{2}_{b,+},
\end{aligned}
\end{eqnarray}
where
\begin{eqnarray*}
\begin{aligned}
P^{2}_{a,-}&=\frac{9}{4}\rho_{44}\rho_{55}+\frac{3}{4}(\rho_{11}\rho_{44}+\rho_{33}\rho_{44}+\rho_{44}\rho_{77}\\
&+\rho_{22}\rho_{55}+\rho_{55}\rho_{66}+\rho_{55}\rho_{88})+\frac{1}{4}(\rho_{11}\rho_{22}\\
&+\rho_{11}\rho_{66}+\rho_{11}\rho_{88}+\rho_{22}\rho_{33}+\rho_{33}\rho_{66}\\
&+\rho_{33}\rho_{88}+\rho_{22}\rho_{77}+\rho_{66}\rho_{77}+\rho_{77}\rho_{88}),
\end{aligned}
\end{eqnarray*}
\begin{eqnarray*}
\begin{aligned}
P^{2}_{a,+}&=\frac{9}{4}\rho_{33}\rho_{66}+\frac{3}{4}(\rho_{22}\rho_{33}+\rho_{33}\rho_{44}+\rho_{33}\rho_{88}\\
&+\rho_{11}\rho_{66}+\rho_{55}\rho_{66}+\rho_{66}\rho_{77})+\frac{1}{4}(\rho_{11}\rho_{22}\\
&+\rho_{11}\rho_{44}+\rho_{11}\rho_{88}+\rho_{22}\rho_{55}+\rho_{44}\rho_{55}\\
&+\rho_{55}\rho_{88}+\rho_{22}\rho_{77}+\rho_{44}\rho_{77}+\rho_{77}\rho_{88}),
\end{aligned}
\end{eqnarray*}
\begin{eqnarray*}
\begin{aligned}
P^{2}_{b,-}&=\frac{9}{4}\rho_{22}\rho_{77}+\frac{3}{4}(\rho_{11}\rho_{22}+\rho_{22}\rho_{33}+\rho_{22}\rho_{55}\\
&+\rho_{44}\rho_{77}+\rho_{66}\rho_{77}+\rho_{77}\rho_{88})+\frac{1}{4}(\rho_{11}\rho_{44}\\
&+\rho_{11}\rho_{66}+\rho_{11}\rho_{88}+\rho_{33}\rho_{44}+\rho_{33}\rho_{66}\\
&+\rho_{33}\rho_{88}+\rho_{44}\rho_{55}+\rho_{55}\rho_{66}+\rho_{55}\rho_{88}),
\end{aligned}
\end{eqnarray*}
\begin{eqnarray}
\begin{aligned}
\label{Pa-d2}
P^{2}_{b,+}&=\frac{9}{4}\rho_{11}\rho_{88}+\frac{3}{4}(\rho_{11}\rho_{22}+\rho_{11}\rho_{44}+\rho_{11}\rho_{66}\\
&+\rho_{33}\rho_{88}+\rho_{55}\rho_{88}+\rho_{77}\rho_{88})+\frac{1}{4}(\rho_{22}\rho_{33}\\
&+\rho_{33}\rho_{44}+\rho_{33}\rho_{66}+\rho_{22}\rho_{55}+\rho_{44}\rho_{55}\\
&+\rho_{55}\rho_{66}+\rho_{22}\rho_{77}+\rho_{44}\rho_{77}+\rho_{66}\rho_{77}).
\end{aligned}
\end{eqnarray}

Consequently, according to Eqs.(\ref{Pa})-(\ref{Pa-d2}), the steering from C to AB is given by
\begin{eqnarray}
\begin{aligned}
\label{SAB-C}
S^{C\rightarrow AB}=&max\left\{0, 4(|\rho_{18}|^{2}-P^{1}_{a,-}), 4(|\rho_{27}|^{2}-P^{1}_{a,+}),\right.\\
&\left.4(|\rho_{36}|^{2}-P^{1}_{b,-}), 4(|\rho_{45}|^{2}-P^{1}_{b,+})\right\}
\end{aligned}
\end{eqnarray}
and the steering from AB to C describes
\begin{eqnarray}
\begin{aligned}
\label{S2AB-C}
S^{AB\rightarrow C}=&max\left\{0, \frac{16}{3}(|\rho_{18}|^{2}-P^{2}_{a,-}), \frac{16}{3}(|\rho_{27}|^{2}-P^{2}_{a,+}),\right.\\
&\left.\frac{16}{3}(|\rho_{36}|^{2}-P^{2}_{b,-}), \frac{16}{3}(|\rho_{45}|^{2}-P^{2}_{b,+})\right\}.
\end{aligned}
\end{eqnarray}
Herein, the coefficient $4$ and $\frac{16}{3}$ are set to guarantee that the steering of the maximally entangled state reaches $1$.

By the same token, we are able to derive the steering from A to BC, BC to A, B to CA and CA to B as
\begin{eqnarray}
\begin{aligned}
\label{SBC-A}
S^{A\rightarrow BC}=&max\left\{0, 4(|\rho_{18}|^{2}-T^{1}_{a,-}), 4(|\rho_{27}|^{2}-T^{1}_{b,-}),\right.\\
&\left.4(|\rho_{36}|^{2}-T^{1}_{b,+}), 4(|\rho_{45}|^{2}-T^{1}_{a,+})\right\},
\end{aligned}
\end{eqnarray}
\begin{eqnarray}
\begin{aligned}
\label{S2BC-A}
S^{BC\rightarrow A}=&max\left\{0, \frac{16}{3}(|\rho_{18}|^{2}-T^{2}_{a,-}), \frac{16}{3}(|\rho_{27}|^{2}-T^{2}_{b,-}),\right.\\
&\left.\frac{16}{3}(|\rho_{36}|^{2}-T^{2}_{b,+}), \frac{16}{3}(|\rho_{45}|^{2}-T^{2}_{a,+})\right\},
\end{aligned}
\end{eqnarray}
\begin{eqnarray}
\begin{aligned}
\label{SCA-B}
S^{B\rightarrow CA}=&max\left\{0, 4(|\rho_{18}|^{2}-Q^{1}_{a,+}), 4(|\rho_{27}|^{2}-Q^{1}_{b,+}),\right.\\
&\left.4(|\rho_{36}|^{2}-Q^{1}_{a,-}), 4(|\rho_{45}|^{2}-Q^{1}_{b,-})\right\},
\end{aligned}
\end{eqnarray}
\begin{eqnarray}
\begin{aligned}
\label{S2CA-B}
S^{CA\rightarrow B}=&max\left\{0, \frac{16}{3}(|\rho_{18}|^{2}-Q^{2}_{a,+}), \frac{16}{3}(|\rho_{27}|^{2}-Q^{2}_{b,+}),\right.\\
&\left.\frac{16}{3}(|\rho_{36}|^{2}-Q^{2}_{a,-}), \frac{16}{3}(|\rho_{45}|^{2}-Q^{2}_{b,-})\right\}.
\end{aligned}
\end{eqnarray}
The explicit form of $T^{1}_{a,\pm}$, $T^{1}_{b,\pm}$, $T^{2}_{a,\pm}$, $T^{2}_{b,\pm}$, $Q^{1}_{a,\pm}$, $Q^{1}_{b,\pm}$, $Q^{2}_{a,\pm}$, $Q^{2}_{b,\pm}$ and detailed discussion, please refer to Appendix A and B.

In practice, quantum steering can be categorized into three distinct types: (i) no-way steering, where the quantum state exhibits zero steerability in all directions;
(ii) two-way steering, where the quantum state is steerable in both reciprocal directions; (iii) one-way steering, where the quantum state is steerable exclusively in one direction. The final scenario exemplifies the inherent asymmetry in quantum steering. To quantify this directional bias, the steering asymmetry between optical modes I,J and mode K is defined as
\begin{eqnarray}
\begin{aligned}
\label{asymmetry}
S^{\Delta}_{IJ|K}=\left|S^{K\rightarrow IJ}-S^{IJ\rightarrow K}\right|.
\end{aligned}
\end{eqnarray}

\section{Tripartite Quantum Steering in the Schwarzschild Black Hole}
Now, we consider the tripartite quantum steering within the Schwarzschild spacetime.
Assume that particles Alice, Bob, and Charlie initially share the following GHZ-state at the asymptotic flatness of the Schwarzschild black hole, which is able to be expressed as
\begin{eqnarray}
\begin{aligned}
\label{GHZ}
|\phi\rangle_{ABC}=\alpha|000\rangle+\sqrt{1-\alpha^{2}}|111\rangle,
\end{aligned}
\end{eqnarray}
where $0<\alpha<1$. Subsequently, we designate that Alice remains within the asymptotically flat region, while Bob and Charlie meander in the vicinity of the event horizon of the black hole. Thereby, from Eq.(\ref{eqK}), Eq.(\ref{GHZ}) can be recast as
\begin{eqnarray}
\begin{aligned}
\label{GHZ-H}
|\phi\rangle_{ABbCc}&=\frac{\alpha}{e^{-\frac{\omega}{T}}+1}|00000\rangle+\frac{\alpha}{e^{\frac{\omega}{T}}+1}|01111\rangle\\
&+\frac{\alpha}{\sqrt{e^{-\frac{\omega}{T}}+e^{\frac{\omega}{T}}+2}}(|00011\rangle+|01100\rangle)\\
&+\sqrt{1-\alpha^{2}}|11111\rangle,
\end{aligned}
\end{eqnarray}
where $\alpha=(e^{-\frac{\omega}{T}}+1)^{-\frac{1}{2}}$ and $\beta=(e^{\frac{\omega}{T}}+1)^{-\frac{1}{2}}$, $B$ and $C$ denote fermionic particles localized outside the event horizon of a black hole, while $b$ and $c$ correspond to antifermionic particles within the event horizon. Note that Anti-Bob and Anti-Charlie correspond to the associated antiparticles situated within the event horizon.

Generally, Alice observes fermion $A$, while Bob and Charlie monitor fermions $B$ and $C$ outside the black hole horizon. Correspondingly, antifermions $b$ and $c$ are observed by Anti-Bob and Anti-Charlie inside the horizon. Due to causal isolation between the black hole interior and exterior, Alice, Bob, and Charlie cannot access quantum information within the event horizon. Therefore, we classify exterior fermionic modes $(A, B, C)$ as accessible and interior antifermionic modes $(b, c)$ as inaccessible.

We now explore all possible scenarios, encompassing both physically accessible and inaccessible modes. Specifically, these include three scenarios: (1) three physically accessible modes, (2) two physically accessible modes, and (3) one physically accessible mode.

\subsection{A. Three physically accessible modes}
We initially examine the scenario involving three physically accessible modes, namely, the fermionic particles $A$, $B$, and $C$. Tracing out the inaccessible modes $b$ and $c$ in Eq.(\ref{GHZ-H}), we obtain the reduced state
\begin{eqnarray}
\begin{aligned}
\label{H-ABC}
\rho^{H}_{ABC}&=\frac{\alpha^{2}}{(e^{-\frac{\omega}{T}}+1)^{2}}|000\rangle\langle000|+\frac{\alpha^{2}}{(e^{\frac{\omega}{T}}+1)^{2}}|011\rangle\langle011|\\
&+\frac{\alpha^{2}}{e^{-\frac{\omega}{T}}+e^{\frac{\omega}{T}}+2}(|001\rangle\langle001|+|010\rangle\langle010|)\\
&+\frac{\alpha\sqrt{1-\alpha^{2}}}{e^{-\frac{\omega}{T}}+1}(|000\rangle\langle111|+|111\rangle\langle000|)\\
&+(1-\alpha^{2})|111\rangle\langle111|.
\end{aligned}
\end{eqnarray}
The density matrix corresponding to the state $\rho^{H}_{ABC}$ takes the form as
\begin{eqnarray}
\label{rhoHABC}
\rho^{H}_{ABC}=
\left (
\begin{array}{cccccccc}
a_{11}      &0               &0                  &0                 &0                &0                     &0                       &a_{18}\\
0           &a_{22}          &0                  &0                 &0                &0                     &0                       &0        \\
0           &0               &a_{33}             &0                 &0                &0                     &0                       &0        \\
0           &0               &0                  &a_{44}            &0                &0                     &0                       &0        \\
0           &0               &0                  &0                 &0                &0                     &0                       &0        \\
0           &0               &0                  &0                 &0                &0                     &0                       &0        \\
0           &0               &0                  &0                 &0                &0                     &0                       &0        \\
a_{81}      &0               &0                  &0                 &0                &0                     &0                       &a_{88}
\end{array}
\right ),
\end{eqnarray}
where $a_{11}=\frac{\alpha^{2}}{(e^{-\frac{\omega}{T}}+1)^{2}}$, $a_{22}=a_{33}=\frac{\alpha^{2}}{e^{-\frac{\omega}{T}}+e^{\frac{\omega}{T}}+2}$, $a_{44}=\frac{\alpha^{2}}{(e^{\frac{\omega}{T}}+1)^{2}}$, $a_{88}=(1-\alpha^{2})$, $a_{18}=a_{81}=\frac{\alpha\sqrt{1-\alpha^{2}}}{e^{-\frac{\omega}{T}}+1}$.

By utilizing the Eqs.(\ref{SAB-C})-(\ref{S2CA-B}), we derive the quantum steering from C to AB, AB to C, A to BC, BC to A, B to CA and CA to B as
\begin{eqnarray}
\begin{aligned}
\label{SHC-AB}
S^{C\rightarrow AB}_{H}=&max\left\{0, 4|a_{18}|^{2}\right\},
\end{aligned}
\end{eqnarray}
\begin{eqnarray}
\begin{aligned}
\label{SHAB-C}
S^{AB\rightarrow C}_{H}=&max\left\{0, \frac{16}{3}\left[|a_{18}|^{2}-(a_{11}+a_{33})\right.\right.\\
&\left.\left.\cdot\left(\frac{1}{4}a_{22}+\frac{3}{4}a_{44}+\frac{1}{4}a_{88}\right)\right]\right\},
\end{aligned}
\end{eqnarray}

\begin{eqnarray}
\begin{aligned}
\label{SHA-BC}
S^{A\rightarrow BC}_{H}=&max\left\{0, 4\left(|a_{18}|^{2}-\frac{1}{2}a_{22}a_{33}\right)\right\},
\end{aligned}
\end{eqnarray}
\begin{eqnarray}
\begin{aligned}
\label{SHBC-A}
S^{BC\rightarrow A}_{H}=&max\left\{0, \frac{16}{3}\left[|a_{18}|^{2}-\frac{3}{4}a_{33}a_{88}\right.\right.\\
&\left.\left.-\frac{1}{4}\left(a_{11}+a_{22}+a_{44}\right)a_{88}\right]\right\},
\end{aligned}
\end{eqnarray}

\begin{eqnarray}
\begin{aligned}
\label{SHB-CA}
S^{B\rightarrow CA}_{H}=&max\left\{0, 4|a_{18}|^{2}\right\},
\end{aligned}
\end{eqnarray}
\begin{eqnarray}
\begin{aligned}
\label{SHCA-B}
S^{CA\rightarrow B}_{H}=&max\left\{0, \frac{16}{3}\left[|a_{18}|^{2}-(a_{33}+a_{44}+a_{88})\right.\right.\\
&\left.\left.\cdot\left(\frac{1}{4}a_{11}+\frac{3}{4}a_{22}\right)\right]\right\}.
\end{aligned}
\end{eqnarray}

Through the aforementioned formulas, we observe that the steerability is associated with the Hawking temperature $T$. Specifically, the Hawking radiation of black hole exerts an influence on the steerability. Therefore, within the framework of Schwarzschild spacetime, the $1\rightarrow2$ steering does not coincide with the $2\rightarrow1$ steering, thereby revealing the presence of steering asymmetry. To quantify the extent of the steering asymmetry, we have derived the following expression
\begin{eqnarray}
\begin{aligned}
&S^{\Delta}_{AB|C}=\left|S^{C\rightarrow AB}_{H}-S^{AB\rightarrow C}_{H}\right|,\\
&S^{\Delta}_{BC|A}=\left|S^{A\rightarrow BC}_{H}-S^{BC\rightarrow A}_{H}\right|,\\
&S^{\Delta}_{CA|B}=\left|S^{B\rightarrow CA}_{H}-S^{CA\rightarrow B}_{H}\right|.\\
\end{aligned}
\end{eqnarray}

\begin{figure*}[htbp]
    \centering
    \begin{minipage}[b]{0.325\textwidth} 
        \includegraphics[width=\linewidth]{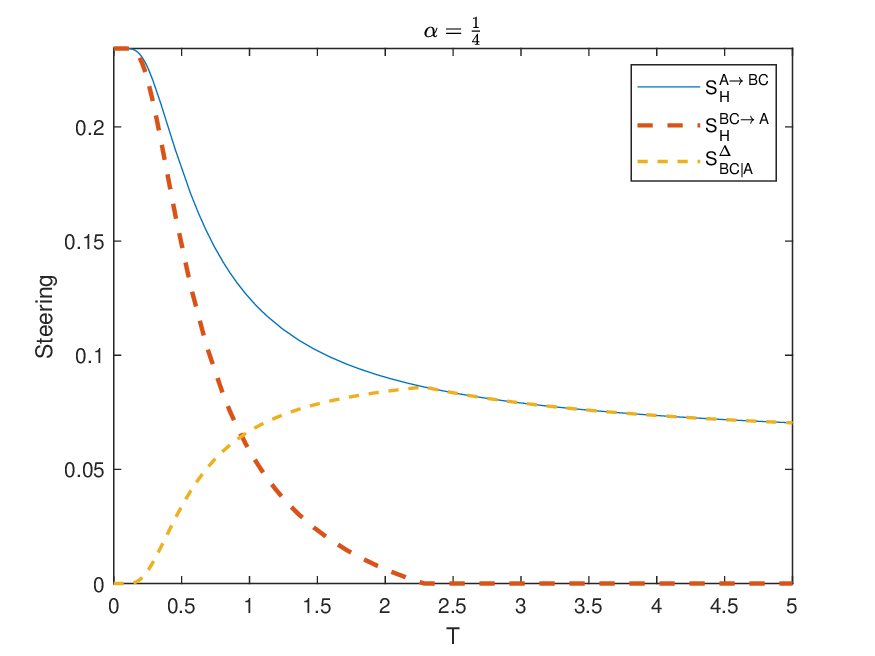} 

    \end{minipage}
    \hfill
    \begin{minipage}[b]{0.325\textwidth}
        \includegraphics[width=\linewidth]{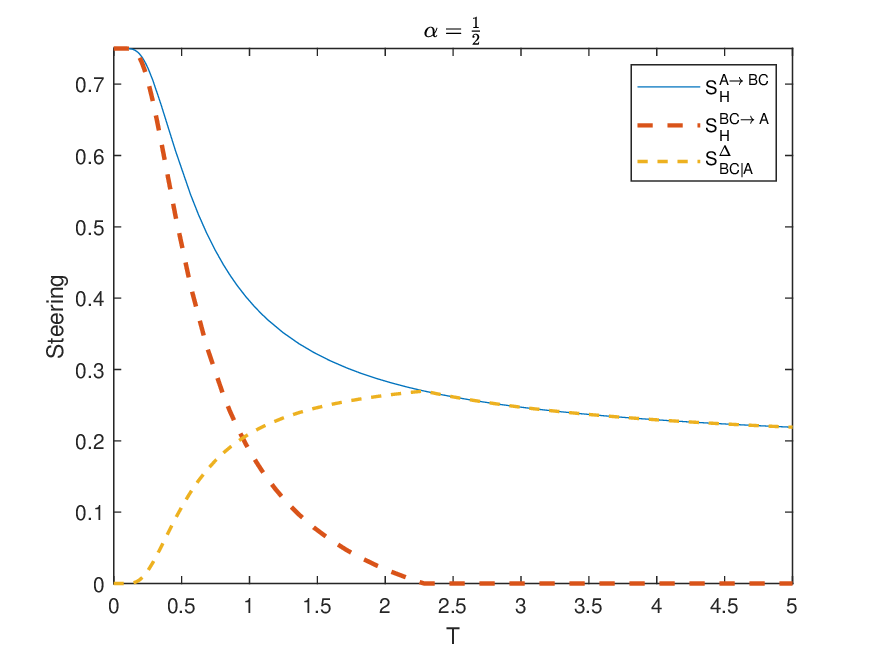}
    \end{minipage}
    \hfill
    \begin{minipage}[b]{0.325\textwidth}
        \includegraphics[width=\linewidth]{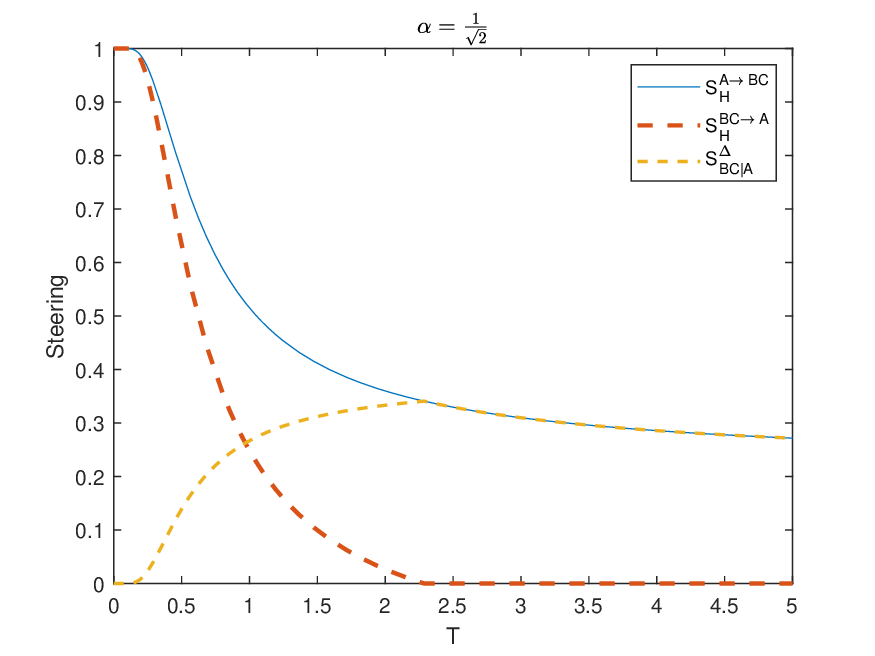}
    \end{minipage}
    \hfill
    \begin{minipage}[b]{0.325\textwidth} 
        \includegraphics[width=\linewidth]{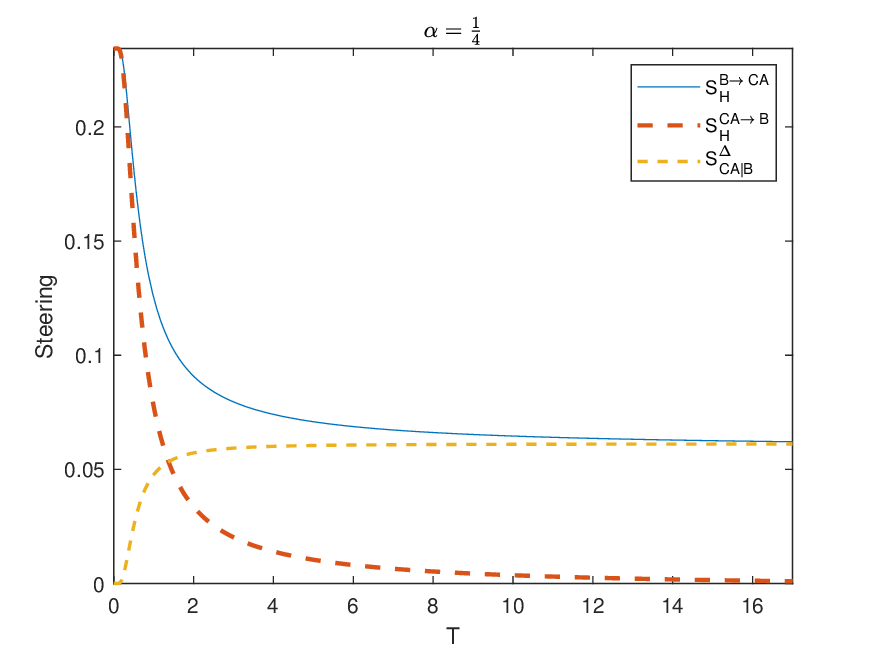} 

    \end{minipage}
    \hfill
    \begin{minipage}[b]{0.325\textwidth}
        \includegraphics[width=\linewidth]{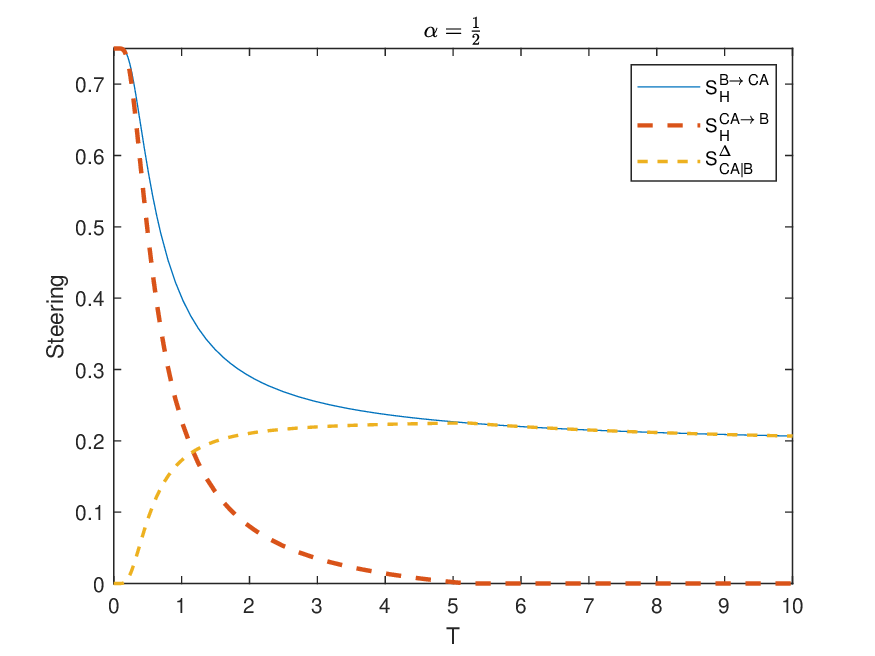}
    \end{minipage}
    \hfill
    \begin{minipage}[b]{0.325\textwidth}
        \includegraphics[width=\linewidth]{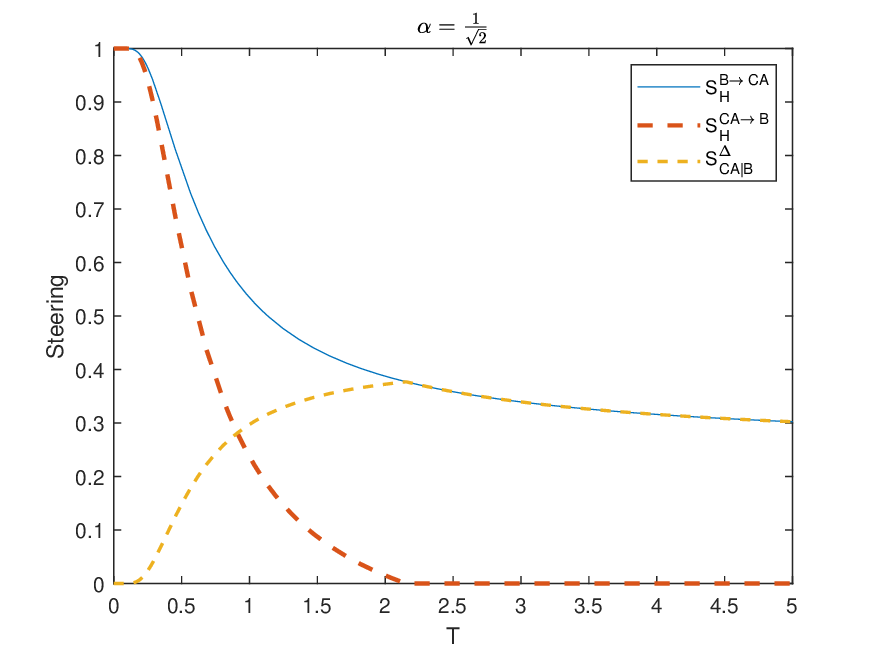}
    \end{minipage}
    \hfill
    \begin{minipage}[b]{0.325\textwidth} 
        \includegraphics[width=\linewidth]{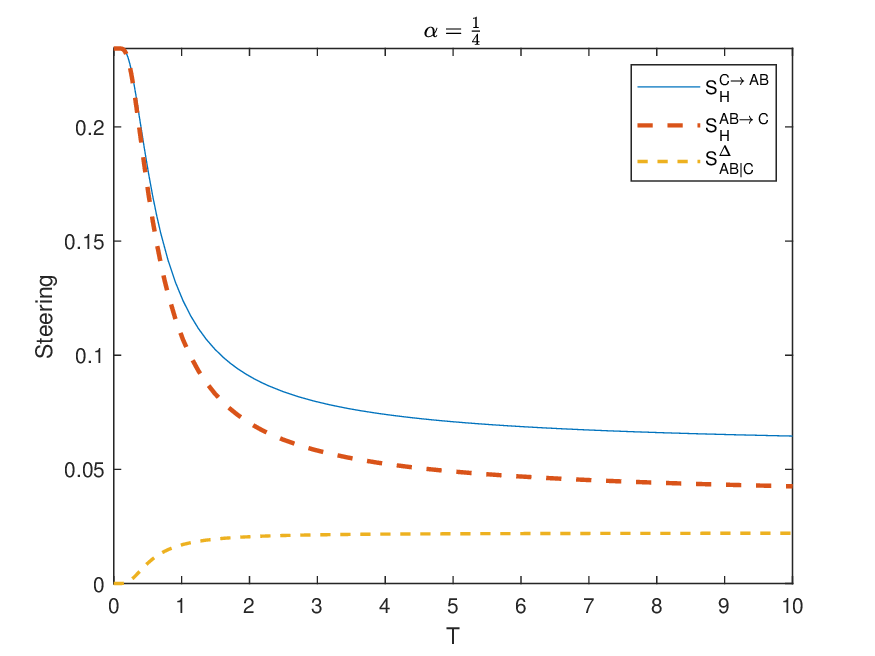} 

    \end{minipage}
    \hfill
    \begin{minipage}[b]{0.325\textwidth}
        \includegraphics[width=\linewidth]{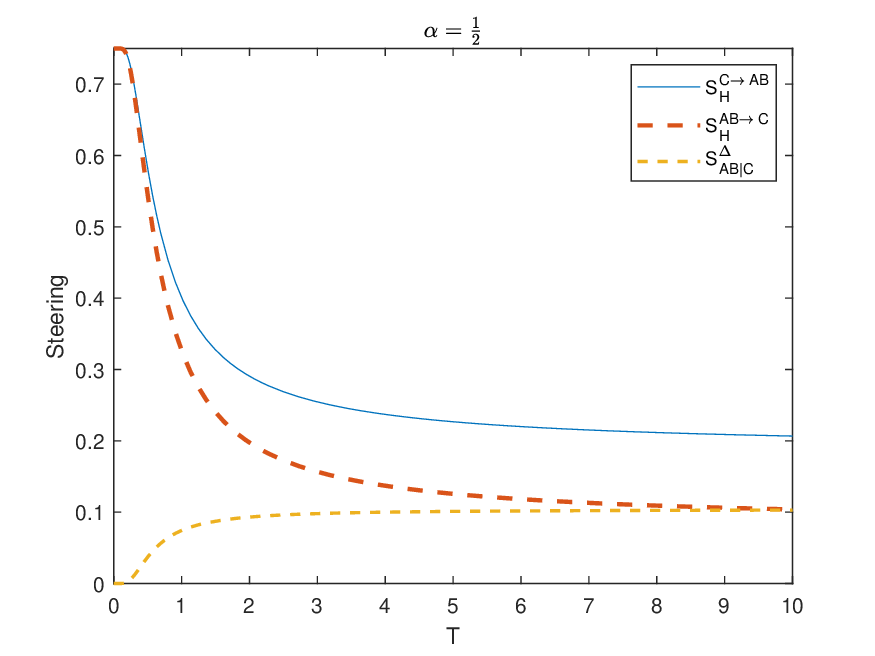}
    \end{minipage}
    \hfill
    \begin{minipage}[b]{0.325\textwidth}
        \includegraphics[width=\linewidth]{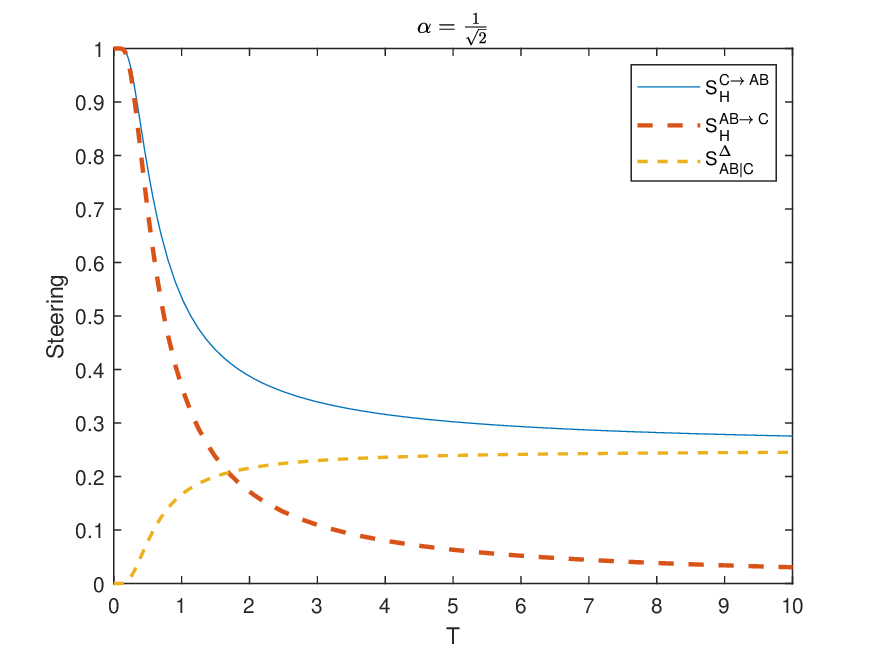}
    \end{minipage}
    \caption{The quantum steering and steering asymmetry of $\rho^{H}_{ABC}$ as functions of the black hole Hawking temperature $T$ with fixed $\omega=1$ and three different values of the parameter $\alpha$: $\alpha=\frac{1}{4}$, $\alpha=\frac{1}{2}$ and $\alpha=\frac{1}{\sqrt{2}}$.}
   \label{FigABC}
\end{figure*}

Fig.\ref{FigABC} illustrates the quantum steering and steering asymmetry of the tripartite state $\rho^{H}_{ABC}$ as functions of the black hole Hawking temperature $T$, with fixed $\omega=1$ and three distinct values of the parameter $\alpha$: $\alpha=\frac{1}{4}$, $\alpha=\frac{1}{2}$ and $\alpha=\frac{1}{\sqrt{2}}$.
The first and second rows of Fig.\ref{FigABC} respectively depict the variations of $S^{A\rightarrow BC}_{H}$ and $S^{BC\rightarrow A}_{H}$ (first row), and $S^{B\rightarrow CA}_{H}$ and $S^{CA\rightarrow B}_{H}$ (second row) with respect to the Hawking temperature $T$. As the Hawking temperature $T$ increases, $S^{A\rightarrow BC}_{H}$, $S^{BC\rightarrow A}_{H}$, $S^{B\rightarrow CA}_{H}$ and $S^{CA\rightarrow B}_{H}$ exhibit a monotonic decrease, indicating that the Hawking effect suppresses quantum steering.
Notably, $S^{BC\rightarrow A}_{H}$ exhibits a ``sudden death'' phenomenon, while $S^{A\rightarrow BC}_{H}$ persists without vanishing entirely as the Hawking temperature $T$ increases. For $S^{CA\rightarrow B}_{H}$, a sudden death phenomenon occurs at $\alpha=\frac{1}{\sqrt{2}}$, while it exhibits a gradual decay to zero when $\alpha=\frac{1}{4}$ and $\alpha=\frac{1}{2}$. In contrast, $S^{B\rightarrow CA}_{H}$ remains non-zero throughout the parameter range.
This indicates a critical quantum transition, where the steering evolves from two-way steering to one-way steering due to the Hawking-induced decoherence.
When the system transitions from two-way steering to one-way steering, the steering asymmetry reaches its maximum. This maximal asymmetry precisely demarcates the phase boundary between two-way steering to one-way steering regimes in black hole spacetime.

For the bottom row of Fig.\ref{FigABC}, the steering $S^{C\rightarrow AB}_{H}$ and $S^{AB\rightarrow C}_{H}$ show a monotonic decrease with rising Hawking temperature $T$, yet their asymptotic values persist above zero, indicating incomplete suppression of quantum correlations.
In addition, from Fig.\ref{FigABC}, it is evident that for a fixed Hawking temperature $T$, the quantum steering  exhibits a monotonic increase with the parameter $\alpha$ and the $1\rightarrow2$ steering intensity  consistently surpasses that the $2\rightarrow1$ steering.

Therefore, we conclude that the Hawking effect of a black hole suppresses quantum steering in the fully physically accessible scenario.
In~\cite{Wu.2025}, the authors conclude that the Unruh effect degrade quantum steering of GHZ states in relativistic motion, which further corroborates our results.
For bipartite states, in~\cite{EPJC.2022}, the authors concluded that the quantum steering first experiences a decline and subsequently approaches a non-zero asymptotic value. This conclusion bears similarity to that derived from the last row in Fig.\ref{FigABC}. In contrast, the $2\rightarrow1$ steering in the first two rows eventually diminishes to zero.

\subsection{B. Two physically accessible modes}
Let us consider the scenario comprising two physically accessible modes: fermionic particles $A$, $B$, and an antifermionic particle $c$. From Eq.(\ref{GHZ-H}), by tracing over the inaccessible modes $b$ and accessible modes $C$ , we have
\begin{eqnarray}
\begin{aligned}
\label{H-ABc}
\rho^{H}_{ABc}&=\frac{\alpha^{2}}{(e^{-\frac{\omega}{T}}+1)^{2}}|000\rangle\langle000|+\frac{\alpha^{2}}{(e^{\frac{\omega}{T}}+1)^{2}}|011\rangle\langle011|\\
&+\frac{\alpha^{2}}{e^{-\frac{\omega}{T}}+e^{\frac{\omega}{T}}+2}(|001\rangle\langle001|+|010\rangle\langle010|)\\
&+\frac{\alpha\sqrt{1-\alpha^{2}}}{\sqrt{e^{-\frac{\omega}{T}}+e^{\frac{\omega}{T}}+2}}(|001\rangle\langle110|+|110\rangle\langle001|)\\
&+(1-\alpha^{2})|110\rangle\langle110|.
\end{aligned}
\end{eqnarray}

The density matrix corresponding to the state $\rho^{H}_{ABc}$ takes the form as

\begin{eqnarray}
\label{rhoHABc}
\rho^{H}_{ABc}=
\left (
\begin{array}{cccccccc}
b_{11}      &0               &0                  &0                 &0                &0                     &0                       &0        \\
0           &b_{22}          &0                  &0                 &0                &0                     &b_{27}                  &0        \\
0           &0               &b_{33}             &0                 &0                &0                     &0                       &0        \\
0           &0               &0                  &b_{44}            &0                &0                     &0                       &0        \\
0           &0               &0                  &0                 &0                &0                     &0                       &0        \\
0           &0               &0                  &0                 &0                &0                     &0                       &0        \\
0           &b_{72}          &0                  &0                 &0                &0                     &b_{77}                  &0        \\
0           &0               &0                  &0                 &0                &0                     &0                       &0
\end{array}
\right ),
\end{eqnarray}
where $b_{11}=\frac{\alpha^{2}}{(e^{-\frac{\omega}{T}}+1)^{2}}$, $b_{22}=b_{33}=\frac{\alpha^{2}}{e^{-\frac{\omega}{T}}+e^{\frac{\omega}{T}}+2}$, $b_{44}=\frac{\alpha^{2}}{(e^{\frac{\omega}{T}}+1)^{2}}$, $b_{77}=(1-\alpha^{2})$, $b_{27}=b_{72}=\frac{\alpha\sqrt{1-\alpha^{2}}}{\sqrt{e^{-\frac{\omega}{T}}+e^{\frac{\omega}{T}}+2}}$.

Then, employing the Eqs.(\ref{SAB-C})-(\ref{S2CA-B}), we deduce the quantum steering from $c$ to $AB$, $AB$ to $c$, $A$ to $Bc$, $Bc$ to $A$, $B$ to $cA$ and $cA$ to $B$ as
\begin{eqnarray}
\begin{aligned}
\label{SHc-AB}
S^{c\rightarrow AB}_{H}=&max\left\{0, 4|b_{27}|^{2}\right\},
\end{aligned}
\end{eqnarray}
\begin{eqnarray}
\begin{aligned}
\label{SHAB-c}
S^{AB\rightarrow c}_{H}=&max\left\{0, \frac{16}{3}\left[|b_{27}|^{2}-(b_{22}+b_{44})\right.\right.\\
&\left.\left.\cdot\left(\frac{1}{4}b_{11}+\frac{3}{4}b_{33}+\frac{1}{4}b_{77}\right)\right]\right\},
\end{aligned}
\end{eqnarray}

\begin{eqnarray}
\begin{aligned}
\label{SHA-Bc}
S^{A\rightarrow Bc}_{H}=&max\left\{0, 4(|b_{27}|^{2}-\frac{1}{2}b_{11}b_{44})\right\},
\end{aligned}
\end{eqnarray}
\begin{eqnarray}
\begin{aligned}
\label{SHBc-A}
S^{Bc\rightarrow A}_{H}=&max\left\{0, \frac{16}{3}\left[|b_{27}|^{2}-\frac{3}{4}b_{44}b_{77}\right.\right.\\
&\left.\left.-\frac{1}{4}\left(b_{11}+b_{22}+b_{33}\right)b_{77}\right]\right\},
\end{aligned}
\end{eqnarray}

\begin{eqnarray}
\begin{aligned}
\label{SHcA-B}
S^{B\rightarrow cA}_{H}=&max\left\{0, 4|b_{27}|^{2}\right\},
\end{aligned}
\end{eqnarray}
\begin{eqnarray}
\begin{aligned}
\label{SHcA-B}
S^{cA\rightarrow B}_{H}=&max\left\{0, \frac{16}{3}\left[|b_{27}|^{2}-(b_{33}+b_{44}+b_{77})\right.\right.\\
&\left.\left.\cdot\left(\frac{3}{4}b_{11}+\frac{1}{4}b_{22}\right)\right]\right\}.
\end{aligned}
\end{eqnarray}

From Eq.(\ref{asymmetry}), we can obtain the steering asymmetry
\begin{eqnarray}
\begin{aligned}
&S^{\Delta}_{AB|c}=\left|S^{c\rightarrow AB}_{H}-S^{AB\rightarrow c}_{H}\right|,\\
&S^{\Delta}_{Bc|A}=\left|S^{A\rightarrow Bc}_{H}-S^{Bc\rightarrow A}_{H}\right|,\\
&S^{\Delta}_{cA|B}=\left|S^{B\rightarrow cA}_{H}-S^{cA\rightarrow B}_{H}\right|.\\
\end{aligned}
\end{eqnarray}

\begin{figure*}[htbp]
    \centering
    \begin{minipage}[b]{0.325\textwidth} 
        \includegraphics[width=\linewidth]{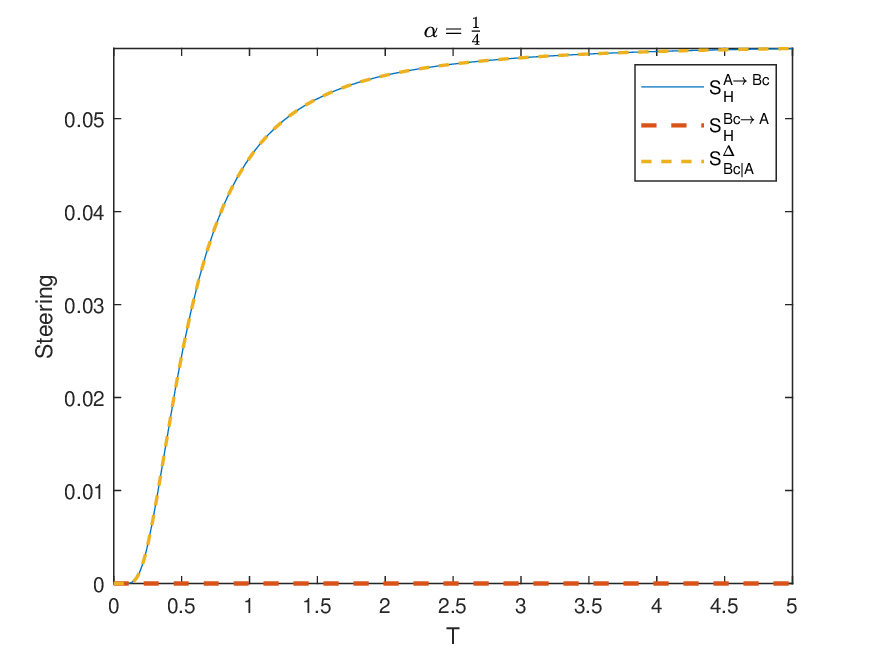} 

    \end{minipage}
    \hfill
    \begin{minipage}[b]{0.325\textwidth}
        \includegraphics[width=\linewidth]{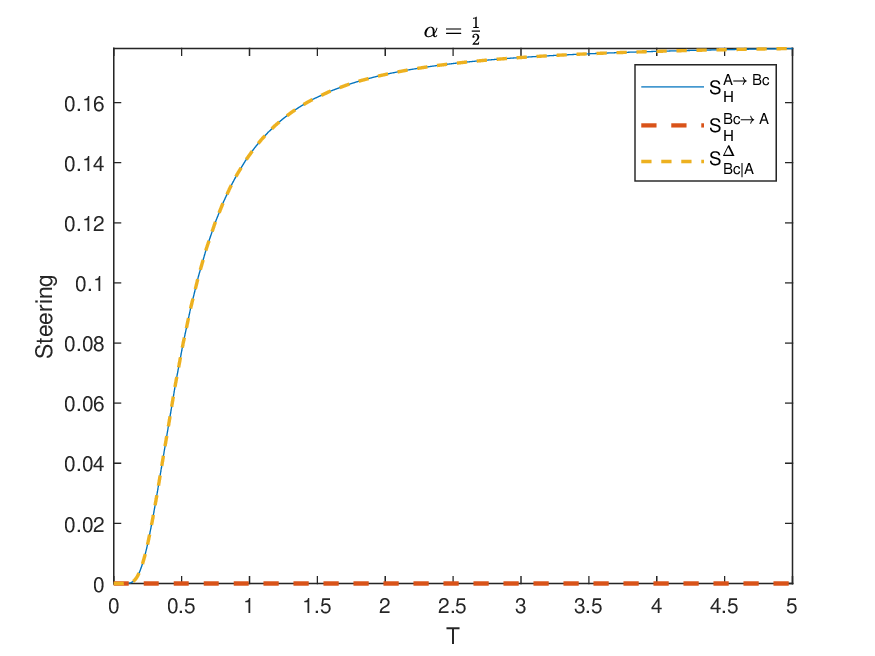}
    \end{minipage}
    \hfill
    \begin{minipage}[b]{0.325\textwidth}
        \includegraphics[width=\linewidth]{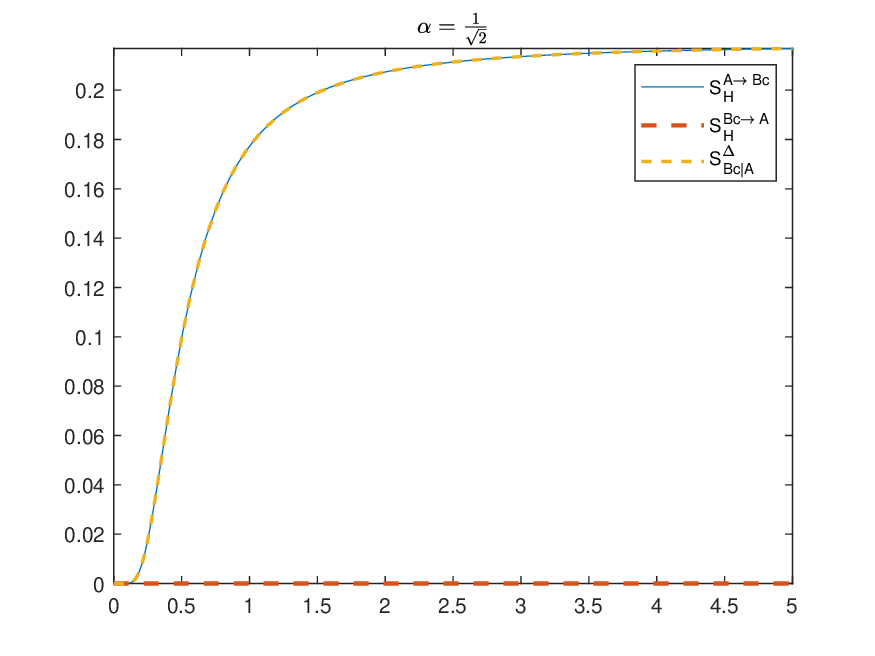}
    \end{minipage}
    \hfill
    \begin{minipage}[b]{0.325\textwidth} 
        \includegraphics[width=\linewidth]{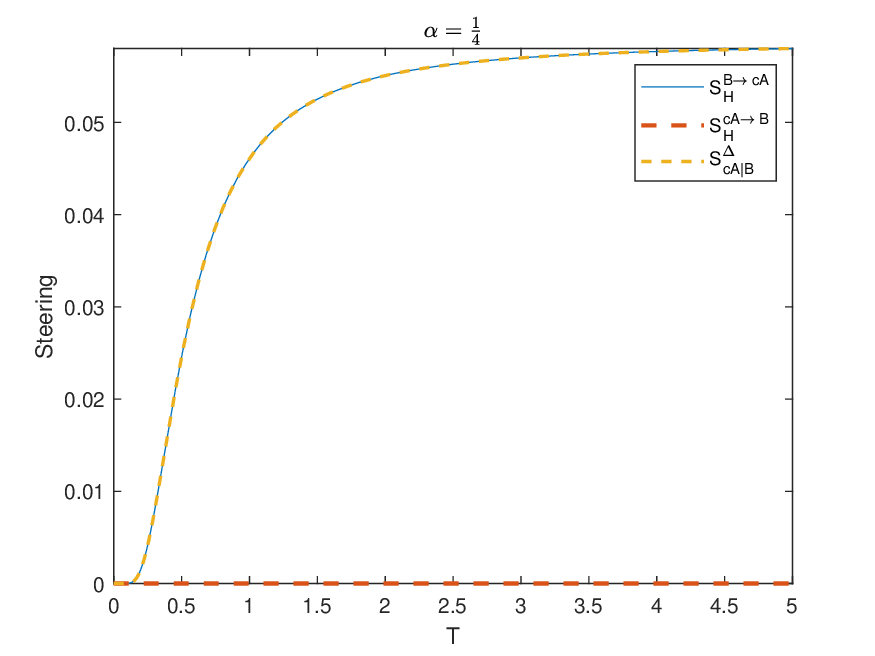} 

    \end{minipage}
    \hfill
    \begin{minipage}[b]{0.325\textwidth}
        \includegraphics[width=\linewidth]{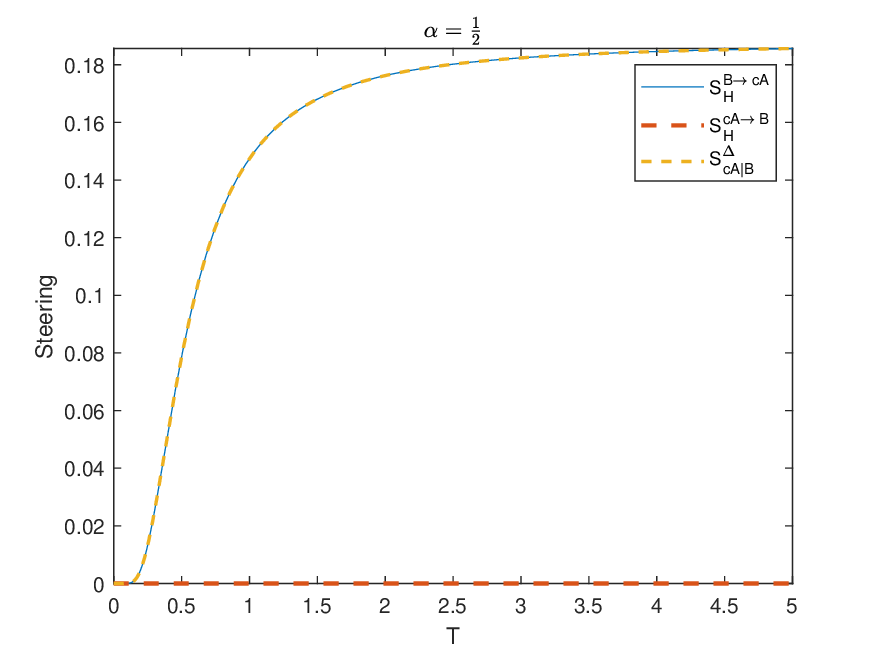}
    \end{minipage}
    \hfill
    \begin{minipage}[b]{0.325\textwidth}
        \includegraphics[width=\linewidth]{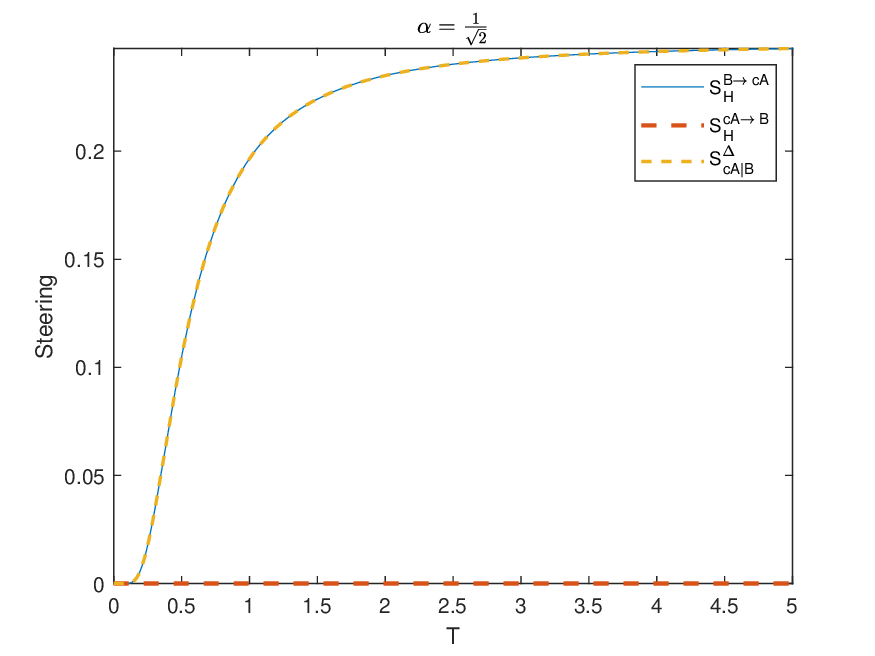}
    \end{minipage}
    \hfill
    \begin{minipage}[b]{0.325\textwidth} 
        \includegraphics[width=\linewidth]{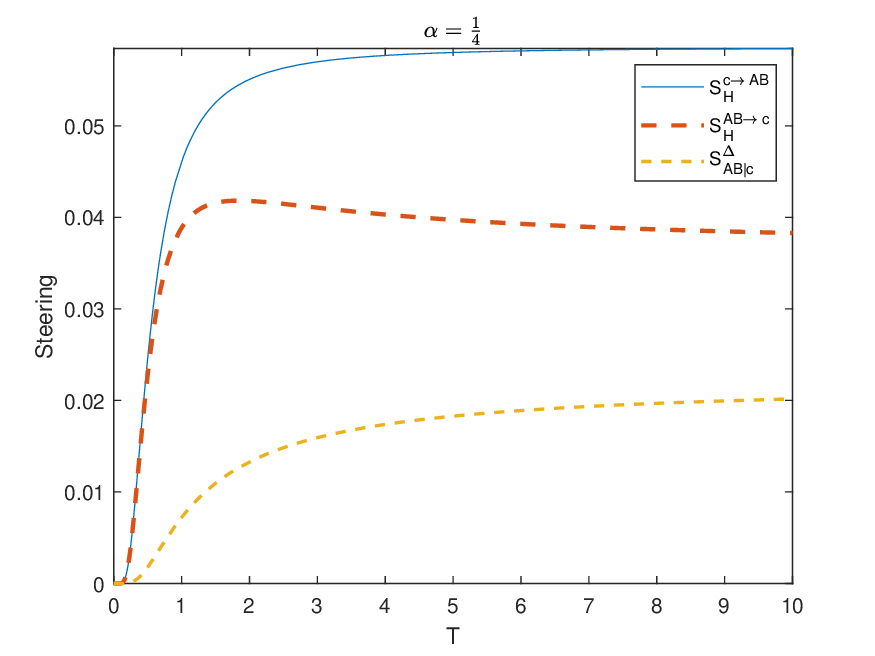} 

    \end{minipage}
    \hfill
    \begin{minipage}[b]{0.325\textwidth}
        \includegraphics[width=\linewidth]{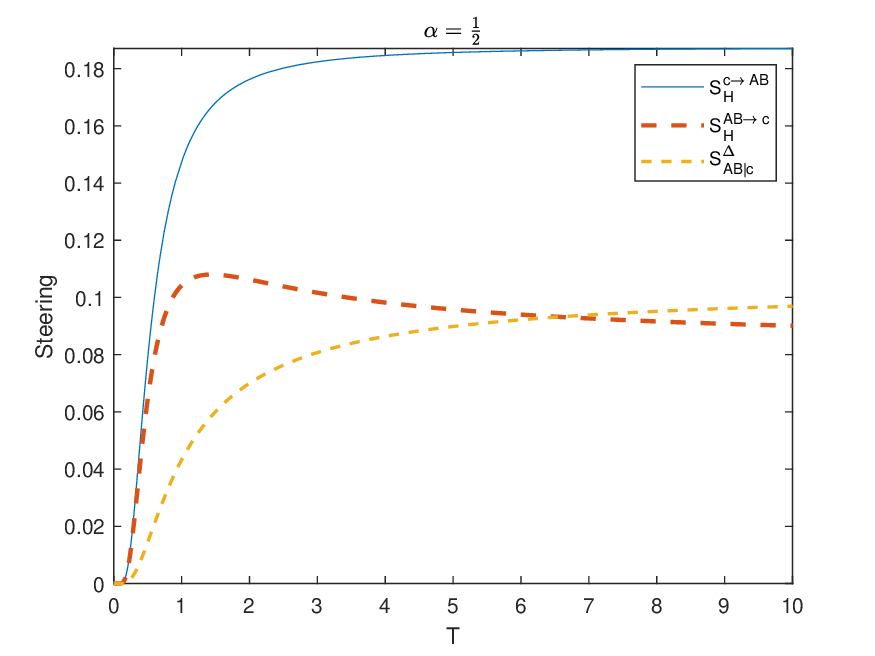}
    \end{minipage}
    \hfill
    \begin{minipage}[b]{0.325\textwidth}
        \includegraphics[width=\linewidth]{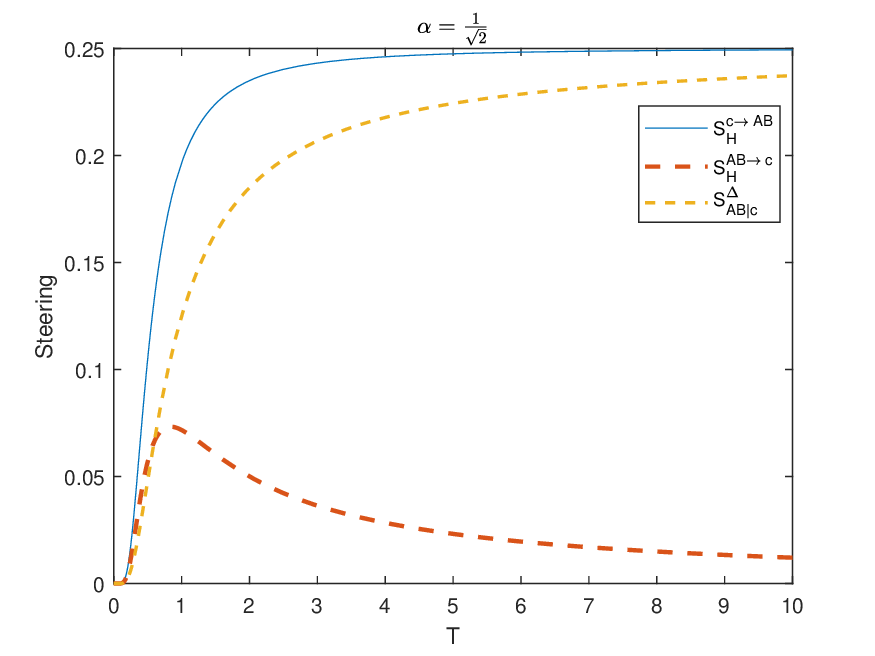}
    \end{minipage}
    \caption{The quantum steering and steering asymmetry of $\rho^{H}_{ABc}$ as functions of the black hole Hawking temperature $T$ with fixed $\omega=1$ and three different values of the parameter $\alpha$: $\alpha=\frac{1}{4}$, $\alpha=\frac{1}{2}$ and $\alpha=\frac{1}{\sqrt{2}}$.}
   \label{FigABc}
\end{figure*}
\begin{figure*}[htbp]
    \centering
    \begin{minipage}[b]{0.325\textwidth} 
        \includegraphics[width=\linewidth]{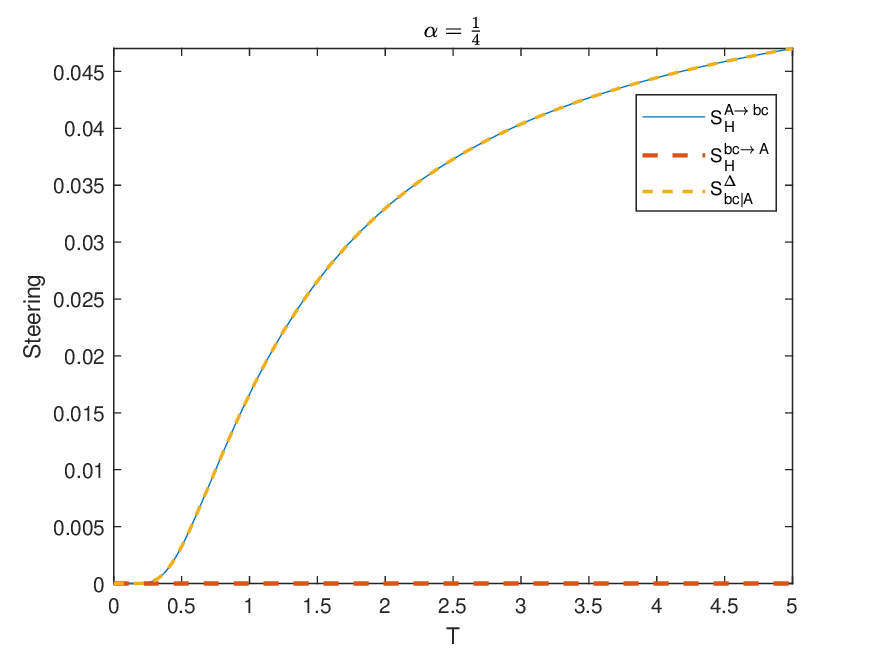} 

    \end{minipage}
    \hfill
    \begin{minipage}[b]{0.325\textwidth}
        \includegraphics[width=\linewidth]{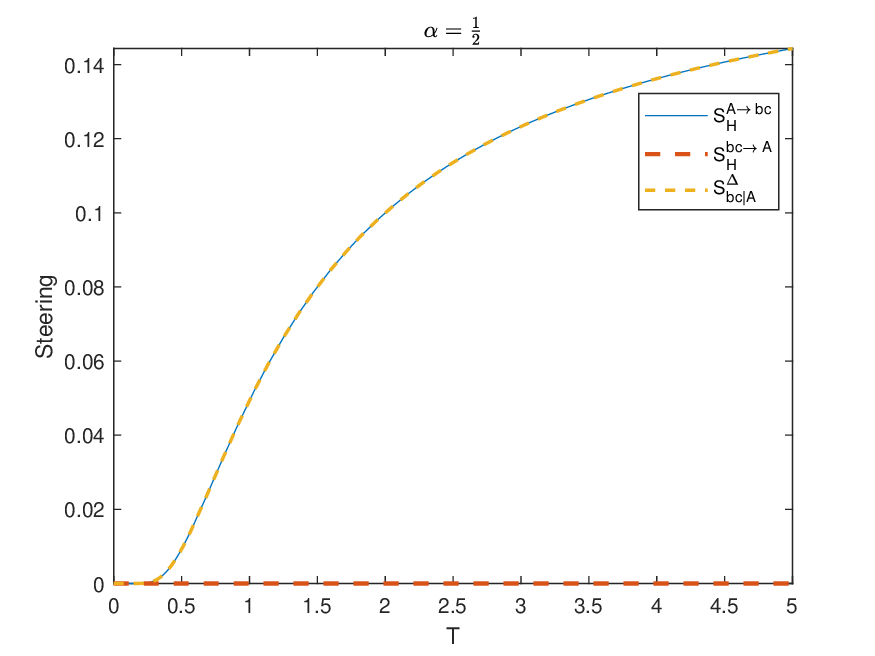}
    \end{minipage}
    \hfill
    \begin{minipage}[b]{0.325\textwidth}
        \includegraphics[width=\linewidth]{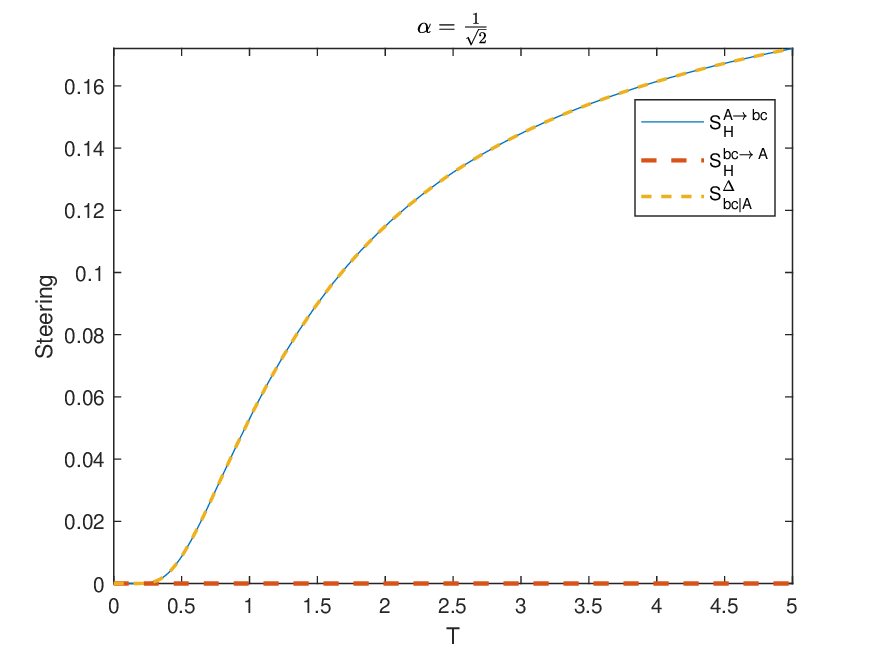}
    \end{minipage}
    \hfill
    \begin{minipage}[b]{0.325\textwidth} 
        \includegraphics[width=\linewidth]{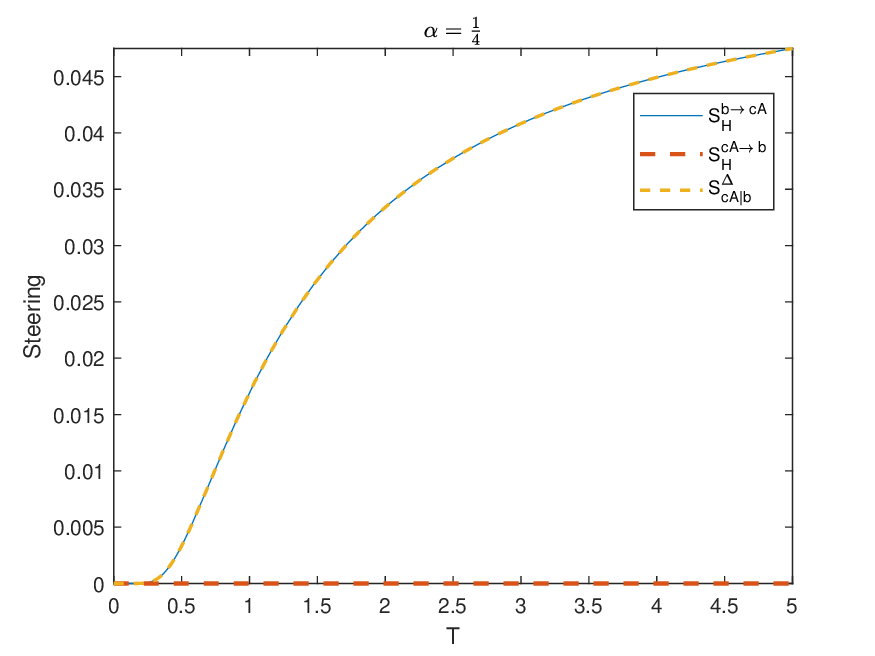} 

    \end{minipage}
    \hfill
    \begin{minipage}[b]{0.325\textwidth}
        \includegraphics[width=\linewidth]{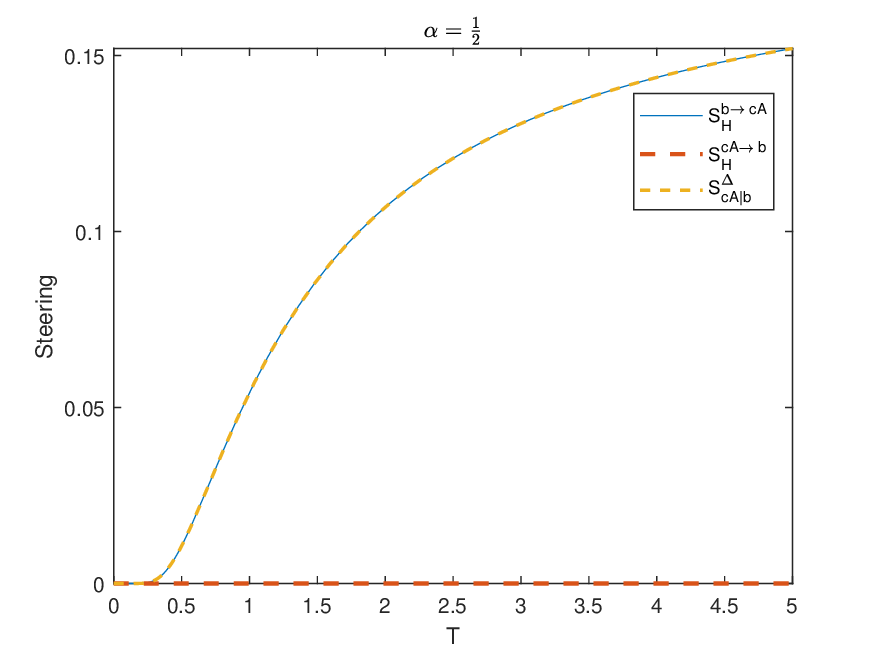}
    \end{minipage}
    \hfill
    \begin{minipage}[b]{0.325\textwidth}
        \includegraphics[width=\linewidth]{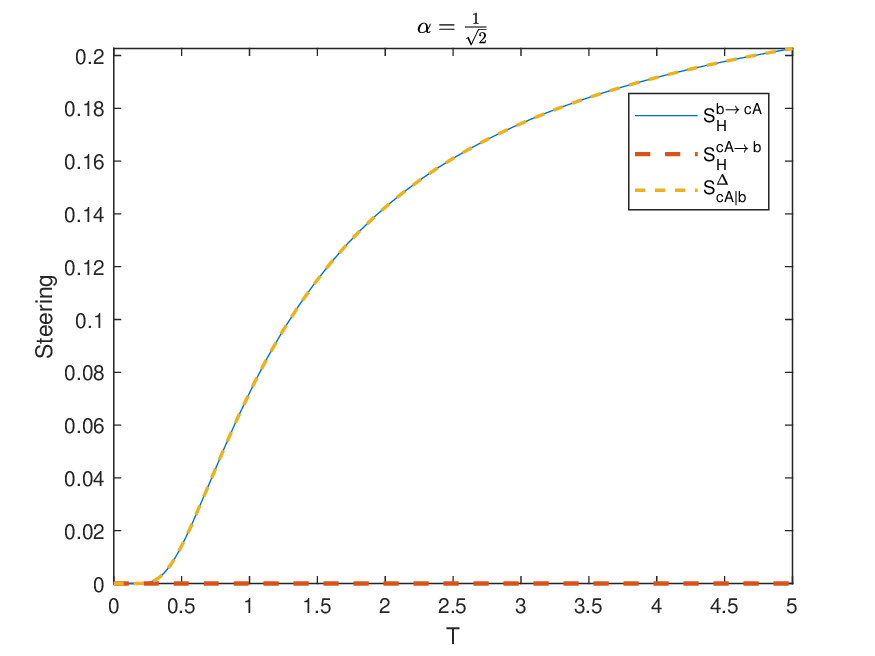}
    \end{minipage}
    \hfill
    \begin{minipage}[b]{0.325\textwidth} 
        \includegraphics[width=\linewidth]{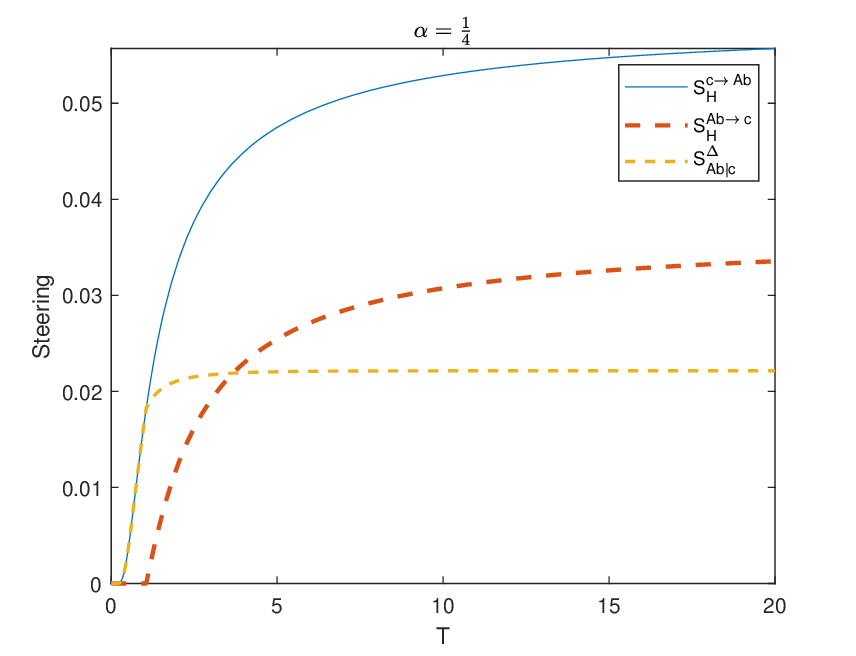} 

    \end{minipage}
    \hfill
    \begin{minipage}[b]{0.325\textwidth}
        \includegraphics[width=\linewidth]{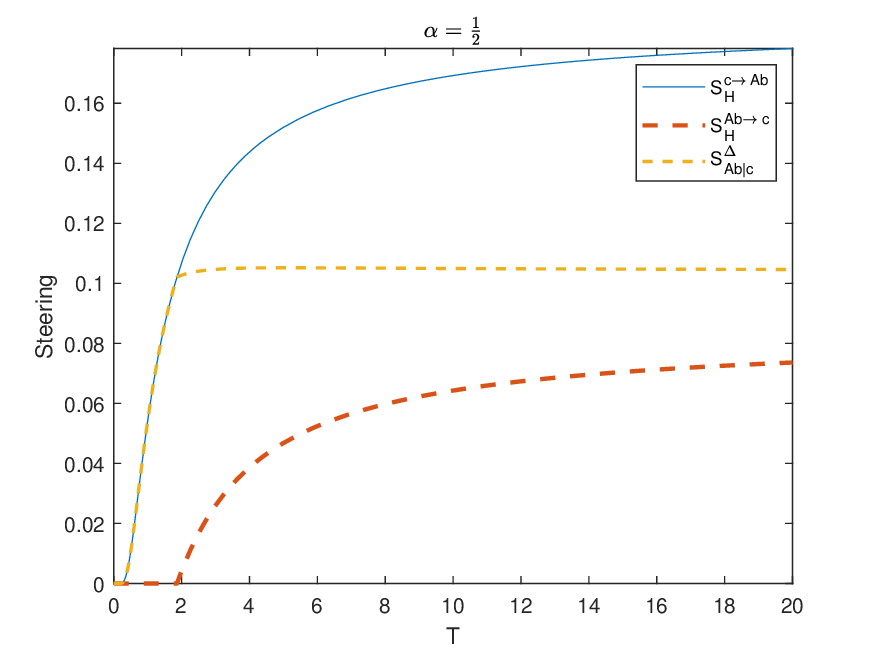}
    \end{minipage}
    \hfill
    \begin{minipage}[b]{0.325\textwidth}
        \includegraphics[width=\linewidth]{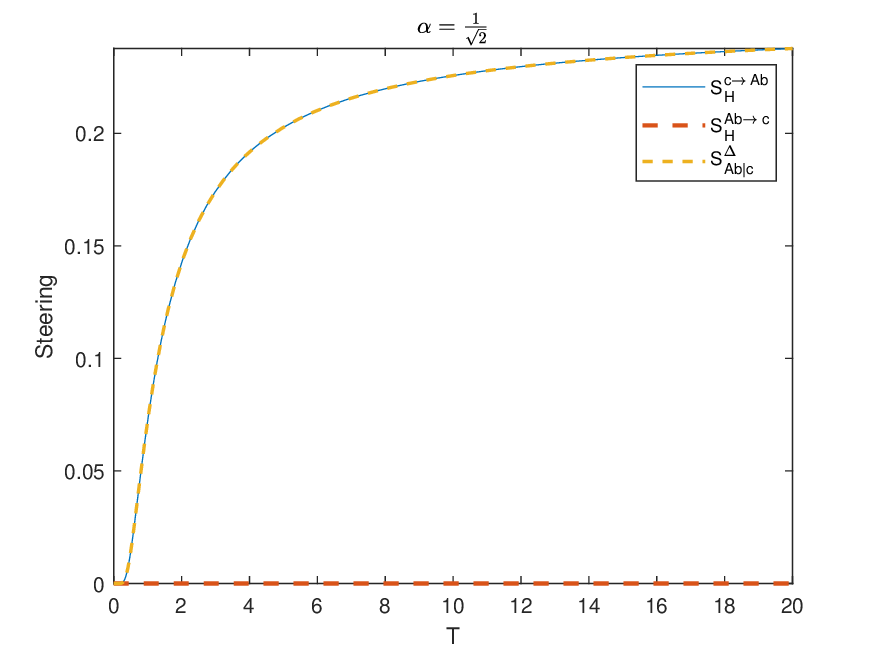}
    \end{minipage}
    \caption{The quantum steering and steering asymmetry of $\rho^{H}_{Abc}$ as functions of the black hole Hawking temperature $T$ with fixed $\omega=1$ and three different values of the parameter $\alpha$: $\alpha=\frac{1}{4}$, $\alpha=\frac{1}{2}$ and $\alpha=\frac{1}{\sqrt{2}}$.}
   \label{FigAbc}
\end{figure*}

In Fig.\ref{FigABc}, we plot the quantum steering and its asymmetry for the tripartite state $\rho^{H}_{ABc}$ versus the black hole Hawking temperature $T$, keeping $\omega=1$ constant while varying $\alpha$ among three representative values: $\frac{1}{4}$, $\frac{1}{2}$ and $\frac{1}{\sqrt{2}}$.
The first row of Fig.\ref{FigABc} plots $S^{A\rightarrow Bc}_{H}$ and $S^{Bc\rightarrow A}_{H}$ against the Hawking temperature $T$, while the second row shows the corresponding behavior of $S^{B\rightarrow cA}_{H}$ and $S^{cA\rightarrow B}_{H}$.
With increasing Hawking temperature $T$, $S^{A\rightarrow Bc}_{H}$ and $S^{B\rightarrow cA}_{H}$ exhibit a gradual rise and subsequent saturation, while $S^{Bc\rightarrow A}_{H}$ and $S^{cA\rightarrow B}_{H}$ remain identically zero throughout. Crucially, $S^{\Delta}_{Bc|A}=S^{A\rightarrow Bc}_{H}$ and $S^{\Delta}_{cA|B}=S^{B\rightarrow cA}_{H}$ at the stage. The observed asymmetry in correlation functions confirms that the system undergoes one-way quantum steering under the current experimental conditions.
In the bottom row of Fig.\ref{FigABc}, $S^{c\rightarrow AB}_{H}$ exhibits a monotonic increase with the Hawking temperature $T$, while $S^{AB\rightarrow c}_{H}$ demonstrates a non-monotonic dependence on $T$ , characterized by an initial rise followed by a gradual decline. This implies a transition of the quantum system from  no-way ($T\rightarrow0$) steering to two-way steering. Besides, as illustrated in Fig.\ref{FigABc}, when the Hawking temperature $T$ is held constant, the quantum steering (excluding the $S^{AB\rightarrow c}_{H}$) demonstrates a monotonic growth with respect to the parameter $\alpha$.

Consequently, as depicted in Fig.\ref{FigABc}, we conclude that the Hawking effect exhibits a dual influence on quantum steering: it not only enhances steering $S^{AB\rightarrow c}_{H}$ but also induces its suppression under specific parameter conditions. Meanwhile, for steering other than $S^{AB\rightarrow c}_{H}$, the Hawking effect demonstrates a net strengthening effect.
\subsection{C. One physically accessible mode}
We analyze the last scenario with one physically accessible mode: fermionic particles $A$, and antifermionic particles $b$ and $c$. By tracing out the accessible modes $B$ and $C$ via Eq.(\ref{GHZ-H}), we obtain
\begin{eqnarray}
\begin{aligned}
\label{H-Abc}
\rho^{H}_{Abc}&=\frac{\alpha^{2}}{(e^{-\frac{\omega}{T}}+1)^{2}}|000\rangle\langle000|+\frac{\alpha^{2}}{(e^{\frac{\omega}{T}}+1)^{2}}|011\rangle\langle011|\\
&+\frac{\alpha^{2}}{e^{-\frac{\omega}{T}}+e^{\frac{\omega}{T}}+2}(|001\rangle\langle001|+|010\rangle\langle010|)\\
&+\frac{\alpha\sqrt{1-\alpha^{2}}}{e^{\frac{\omega}{T}}+1}(|011\rangle\langle100|+|100\rangle\langle011|)\\
&+(1-\alpha^{2})|100\rangle\langle100|.
\end{aligned}
\end{eqnarray}

The density matrix corresponding to the state $\rho^{H}_{ABc}$ takes the form as
\begin{eqnarray}
\label{rhoHAbc}
\rho^{H}_{ABc}=
\left (
\begin{array}{cccccccc}
c_{11}      &0               &0                  &0                 &0                &0                     &0                       &0        \\
0           &c_{22}          &0                  &0                 &0                &0                     &0                       &0        \\
0           &0               &c_{33}             &0                 &0                &0                     &0                       &0        \\
0           &0               &0                  &c_{44}            &c_{45}           &0                     &0                       &0        \\
0           &0               &0                  &c_{54}            &c_{55}           &0                     &0                       &0        \\
0           &0               &0                  &0                 &0                &0                     &0                       &0        \\
0           &0               &0                  &0                 &0                &0                     &0                       &0        \\
0           &0               &0                  &0                 &0                &0                     &0                       &0
\end{array}
\right ),
\end{eqnarray}
where $c_{11}=\frac{\alpha^{2}}{(e^{-\frac{\omega}{T}}+1)^{2}}$, $c_{22}=c_{33}=\frac{\alpha^{2}}{e^{-\frac{\omega}{T}}+e^{\frac{\omega}{T}}+2}$, $c_{44}=\frac{\alpha^{2}}{(e^{\frac{\omega}{T}}+1)^{2}}$, $c_{55}=(1-\alpha^{2})$, $c_{45}=c_{54}=\frac{\alpha\sqrt{1-\alpha^{2}}}{e^{\frac{\omega}{T}}+1}$.

According to the Eqs.(\ref{SAB-C})-(\ref{S2CA-B}), we induce the steering from $c$ to $Ab$, from $A$ to $bc$ and from $b$ to $cA$ as
\begin{eqnarray}
\begin{aligned}
\label{SHc-Ab}
S^{c\rightarrow Ab}_{H}=&max\left\{0, 4|c_{45}|^{2}\right\},
\end{aligned}
\end{eqnarray}
\begin{eqnarray}
\begin{aligned}
\label{SHAb-c}
S^{Ab\rightarrow c}_{H}=&max\left\{0, \frac{16}{3}\left[|c_{45}|^{2}-(c_{22}+c_{44})\right.\right.\\
&\left.\left.\cdot\left(\frac{3}{4}c_{11}+\frac{1}{4}c_{33}+\frac{1}{4}c_{55}\right)\right]\right\},
\end{aligned}
\end{eqnarray}

\begin{eqnarray}
\begin{aligned}
\label{SHA-bc}
S^{A\rightarrow bc}_{H}=&max\left\{0, 4(|c_{45}|^{2}-\frac{1}{2}c_{22}c_{33})\right\},
\end{aligned}
\end{eqnarray}
\begin{eqnarray}
\begin{aligned}
\label{SHbc-A}
S^{bc\rightarrow A}_{H}=&max\left\{0, \frac{16}{3}\left[|c_{45}|^{2}-\frac{3}{4}c_{22}c_{55}\right.\right.\\
&\left.\left.-\frac{1}{4}\left(c_{11}+c_{33}+c_{44}\right)c_{55}\right]\right\},
\end{aligned}
\end{eqnarray}

\begin{eqnarray}
\begin{aligned}
\label{SHb-cA}
S^{b\rightarrow cA}_{H}=&max\left\{0, 4|c_{45}|^{2}\right\},
\end{aligned}
\end{eqnarray}
\begin{eqnarray}
\begin{aligned}
\label{SHcA-b}
S^{cA\rightarrow b}_{H}=&max\left\{0, \frac{16}{3}\left[|c_{45}|^{2}-(c_{11}+c_{22}+c_{55})\right.\right.\\
&\left.\left.\cdot\left(\frac{3}{4}c_{33}+\frac{1}{4}c_{44}\right)\right]\right\}.
\end{aligned}
\end{eqnarray}

According to Eq.(\ref{asymmetry}), we have
\begin{eqnarray}
\begin{aligned}
&S^{\Delta}_{AB|c}=\left|S^{c\rightarrow Ab}_{H}-S^{Ab\rightarrow c}_{H}\right|,\\
&S^{\Delta}_{Bc|A}=\left|S^{A\rightarrow bc}_{H}-S^{bc\rightarrow A}_{H}\right|,\\
&S^{\Delta}_{cA|B}=\left|S^{b\rightarrow cA}_{H}-S^{cA\rightarrow b}_{H}\right|.\\
\end{aligned}
\end{eqnarray}

In Fig.\ref{FigAbc}, we present the quantum steering and steering asymmetry for the tripartite state $\rho^{H}_{Abc}$  as functions of the black hole Hawking temperature $T$. The analysis is conducted under fixed $\omega=1$, with $\alpha$ varied across three characteristic values: $\frac{1}{4}$, $\frac{1}{2}$ and $\frac{1}{\sqrt{2}}$.
The first two rows of Fig.\ref{FigAbc} exhibit qualitative agreement with those of Fig.\ref{FigABc} in terms of the observed trends for quantum steering and its asymmetry.
We then proceed to analyze the bottem row of Fig.\ref{FigAbc}, which depicts the variations of quantum steering parameters $S^{c\rightarrow Ab}_{H}$ and $S^{Ab\rightarrow c}_{H}$ with respect to the Hawking temperature $T$. Notably, when $\alpha=\frac{1}{\sqrt{2}}$, the observed trends align with those in the first two rows, whereas distinct behavioral patterns emerge for $\alpha=\frac{1}{4}$ and $\alpha=\frac{1}{2}$.
When $\alpha=\frac{1}{4}$ ($\alpha=\frac{1}{2}$), both $S^{c\rightarrow Ab}_{H}$ and $S^{Ab\rightarrow c}_{H}$  demonstrate a monotonic increase and appear ``sudden birth'' behavior with the Hawking temperature $T$. Notably, for $T\rightarrow0$, both $S^{c\rightarrow Ab}_{H}$ and $S^{Ab\rightarrow c}_{H}$ remain null and as the Hawking temperature $T$ increases, parameter $S^{c\rightarrow Ab}_{H}$ exhibits an earlier onset of growth from zero compared to parameter $S^{Ab\rightarrow c}_{H}$.
In other words, this reveal a sequential phase transition in quantum steering: from no-way steering to one-way steering, and subsequently to two-way steering as the Hawking temperature increases. 

Furthermore, from Fig.\ref{FigAbc} we show that under the scenario of one physically accessible mode, the Hawking effect enhances quantum steering in the tripartite system.
In~\cite{EPJC.2022}, after conducting research on physically inaccessible quantum steering, the authors found that as the Hawking temperature $T$ increases, the quantum steering $S^{A\rightarrow \bar{B}}$ and $S^{\bar{B}\rightarrow A}$ initially grow gradually from zero and exhibit a ``sudden birth'' behavior. This is similar to some of our results.

\section{Conclusion}
We have investigated the impact of Hawking radiation on quantum steering and steering asymmetry in a tripartite system embedded in Schwarzschild black hole.
In the paper, assume a spacetime configuration where Alice occupies an asymptotically flat spacetime region, while Bob and Charlie orbit near the event horizon of a Schwarzschild black hole. Then, we systematically explore all conceivable scenarios, spanning both physically accessible and inaccessible modes, and specifically detail three scenarios: (i) three physically accessible modes, (ii) two physically accessible modes, and (iii) one physically accessible mode.

In the first scenario, the Hawking effect disrupts quantum steering. Moreover, the attainment of maximum steering asymmetry as the system shifts from two-way to one-way steering precisely marks the phase boundary separating the two-way and one-way steering regimes in black hole spacetime.
We have also conducted a comparison between the fermionic tripartite steering investigated in this paper and the previously studied bipartite steering. Previous research has demonstrated that the fermionic bipartite steering does not exhibit the phenomenon of ``sudden death''~\cite{EPJC.2022}, whereas the bosonic steering undergoes ``sudden death'' during its degradation process~\cite{PRD.2016,APB.2018}. Our results indicate that the steering $S^{BC\rightarrow A}_{H}$ and $S^{CA\rightarrow B}_{H}$ manifest the ``sudden death'' phenomenon, while the other four steering configurations do not display such behavior.

In the second scenario, the Hawking effect exerts a dual-faceted impact on quantum steering. Specifically, it not only boosts the steering $S^{AB\rightarrow c}_{H}$
but also triggers its suppression when certain parameter conditions are met. Simultaneously, for steering types other than $S^{AB\rightarrow c}_{H}$, the Hawking effect manifests a net enhancement effect.
In the third scenario, our analysis reveals that the Hawking effect induces a significant enhancement in quantum steering.

Interestingly, across all three considered scenarios, the $1\rightarrow2$ steering strength consistently surpasses that of the $2\rightarrow1$ steering.
Our results have significantly enriched the theoretical framework of tripartite quantum steering within black  hole spacetime. The findings hold considerable importance for advancing the study of quantum steering in curved spacetime, providing new insights and theoretical foundations for further exploration in this field.

\begin{acknowledgments}
This work is supported by the National Natural Science Foundation of China (NSFC) under Grant Nos. 12204137 and 12564048; the Natural Science Foundation of Hainan Province under Grant No. 125RC744 and the Hainan Academician Workstation.
\end{acknowledgments}

\appendix
\section{APPENDIX A: The steering from A to BC and BC to A}
\label{App1}
Since $\rho_{ABC}=\rho_{X}$, we can deduce that
\begin{eqnarray}
\label{rhoBCA}
\rho_{BCA}=
\left (
\begin{array}{cccccccc}
\rho_{11}   &0               &0                  &0                 &0                &0                     &0                       &\rho_{18}\\
0           &\rho_{55}       &0                  &0                 &0                &0                     &\rho_{45}               &0        \\
0           &0               &\rho_{22}          &0                 &0                &\rho_{27}             &0                       &0        \\
0           &0               &0                  &\rho_{66}         &\rho_{36}        &0                     &0                       &0        \\
0           &0               &0                  &\rho_{63}         &\rho_{33}        &0                     &0                       &0        \\
0           &0               &\rho_{72}          &0                 &0                &\rho_{77}             &0                       &0        \\
0           &\rho_{54}       &0                  &0                 &0                &0                     &\rho_{44}               &0        \\
\rho_{81}   &0               &0                  &0                 &0                &0                     &0                       &\rho_{88}
\end{array}
\right ).
\end{eqnarray}

Hence, in accordance with Eq.(\ref{BCA}), we obtain that the steered density matrix $\tau^{1}_{BC|A}$ associated with $\rho_{BCA}$ can be precisely represented as
\begin{widetext}
\begin{eqnarray}
\label{tauBCA}
\tau^{1}_{BC|A}=
\left (
\begin{array}{cccccccc}
\frac{\sqrt{3}}{3}\rho_{11}+g_{1}  &0               &0                  &0                 &0                &0                     &0            &\frac{\sqrt{3}}{3}\rho_{18}\\
0            &\frac{\sqrt{3}}{3}\rho_{55}+g_{1}     &0                  &0                 &0                &0                     &\frac{\sqrt{3}}{3}\rho_{45}           &0\\
0            &0               &\frac{\sqrt{3}}{3}\rho_{22}+g_{2}          &0                 &0                &\frac{\sqrt{3}}{3}\rho_{27}             &0                 &0\\
0            &0               &0            &\frac{\sqrt{3}}{3}\rho_{66}+g_{2}       &\frac{\sqrt{3}}{3}\rho_{36}          &0                     &0                       &0\\
0            &0               &0            &\frac{\sqrt{3}}{3}\rho_{63}         &\frac{\sqrt{3}}{3}\rho_{33}+g_{3}        &0                     &0                       &0\\
0            &0               &\frac{\sqrt{3}}{3}\rho_{72}    &0                 &0                &\frac{\sqrt{3}}{3}\rho_{77}+g_{3}             &0                       &0\\
0            &\frac{\sqrt{3}}{3}\rho_{54}       &0            &0                 &0                &0                  &\frac{\sqrt{3}}{3}\rho_{44}+g_{4}                  &0\\
\frac{\sqrt{3}}{3}\rho_{81}    &0               &0            &0                 &0                &0                  &0                     &\frac{\sqrt{3}}{3}\rho_{88}+g_{4}
\end{array}
\right ),
\end{eqnarray}
\end{widetext}
where $g_{1}=\frac{3-\sqrt{3}}{6}(\rho_{11}+\rho_{55})$, $g_{2}=\frac{3-\sqrt{3}}{6}(\rho_{22}+\rho_{66})$, $g_{3}=\frac{3-\sqrt{3}}{6}(\rho_{33}+\rho_{77})$, and $g_{4}=\frac{3-\sqrt{3}}{6}(\rho_{44}+\rho_{88})$.
The state $\tau^{1}_{BC|A}$ is entangled provided that it satisfies the inequality
\begin{eqnarray}
\begin{aligned}
\label{Ta}
|\rho_{18}|^{2}>T^{1}_{a,-} \quad \quad \quad or \quad \quad \quad |\rho_{27}|^{2}>T^{1}_{b,-},
\end{aligned}
\end{eqnarray}
or
\begin{eqnarray}
\begin{aligned}
\label{Tc}
|\rho_{36}|^{2}>T^{1}_{b,+} \quad \quad \quad or \quad \quad \quad |\rho_{45}|^{2}>T^{1}_{a,+},
\end{aligned}
\end{eqnarray}
where
\begin{eqnarray*}
\begin{aligned}
T^{1}_{a,\pm}=&\frac{2\pm\sqrt{3}}{2}\rho_{22}\rho_{77}+\frac{2\mp\sqrt{3}}{2}\rho_{33}\rho_{66}\\
&+\frac{1}{2}(\rho_{22}\rho_{33}+\rho_{66}\rho_{77}),\\
\end{aligned}
\end{eqnarray*}
\begin{eqnarray}
\begin{aligned}
\label{Ta.d}
T^{1}_{b,\pm}=&\frac{2\pm\sqrt{3}}{2}\rho_{11}\rho_{88}+\frac{2\mp\sqrt{3}}{2}\rho_{44}\rho_{55}\\
&+\frac{1}{2}(\rho_{11}\rho_{44}+\rho_{55}\rho_{88}).\\
\end{aligned}
\end{eqnarray}
Hence, the steering from A to BC is discerned. Following this, we conclude that the steered density matrix $\tau^{2}_{BC|A}$ corresponding to $\rho_{BCA}$ can be exactly expressed as
\begin{widetext}
\begin{eqnarray}
\label{tau2BCA}
\tau^{2}_{BC|A}=
\left (
\begin{array}{cccccccc}
\frac{1}{3}\rho_{11}+h_{1}  &0               &0                  &0                 &0                &0                     &0            &\frac{1}{3}\rho_{18}\\
0            &\frac{1}{3}\rho_{55}+h_{2}     &0                  &0                 &0                &0                     &\frac{1}{3}\rho_{45}           &0\\
0            &0               &\frac{1}{3}\rho_{22}+h_{1}          &0                 &0                &\frac{1}{3}\rho_{27}             &0                 &0\\
0            &0               &0            &\frac{1}{3}\rho_{66}+h_{2}       &\frac{1}{3}\rho_{36}          &0                     &0                       &0\\
0            &0               &0            &\frac{1}{3}\rho_{63}         &\frac{1}{3}\rho_{33}+h_{1}        &0                     &0                       &0\\
0            &0               &\frac{1}{3}\rho_{72}    &0                 &0                &\frac{1}{3}\rho_{77}+h_{2}             &0                       &0\\
0            &\frac{1}{3}\rho_{54}       &0            &0                 &0                &0                  &\frac{1}{3}\rho_{44}+h_{1}                  &0\\
\frac{1}{3}\rho_{81}    &0               &0            &0                 &0                &0                  &0                     &\frac{1}{3}\rho_{88}+h_{2}
\end{array}
\right ),
\end{eqnarray}
\end{widetext}
where $h_{1}=\frac{1}{6}(\rho_{11}+\rho_{22}+\rho_{33}+\rho_{44})$ and $h_{2}=\frac{1}{6}(\rho_{55}+\rho_{66}+\rho_{77}+\rho_{88})$.
The state $\tau^{2}_{BC|A}$ is entangled whenever it satisfies the inequalities
\begin{eqnarray}
\begin{aligned}
\label{Ta2}
|\rho_{18}|^{2}>T^{2}_{a,-} \quad \quad \quad or \quad \quad \quad |\rho_{27}|^{2}>T^{2}_{b,-},
\end{aligned}
\end{eqnarray}
or
\begin{eqnarray}
\begin{aligned}
\label{Tc2}
|\rho_{36}|^{2}>T^{2}_{b,+} \quad \quad \quad or \quad \quad \quad |\rho_{45}|^{2}>T^{2}_{a,+},
\end{aligned}
\end{eqnarray}
where
\begin{eqnarray*}
\begin{aligned}
T^{2}_{a,-}&=\frac{9}{4}\rho_{33}\rho_{66}+\frac{3}{4}(\rho_{11}\rho_{66}+\rho_{22}\rho_{66}+\rho_{44}\rho_{66}\\
&+\rho_{33}\rho_{55}+\rho_{33}\rho_{77}+\rho_{33}\rho_{88})+\frac{1}{4}(\rho_{11}\rho_{55}\\
&+\rho_{11}\rho_{77}+\rho_{11}\rho_{88}+\rho_{22}\rho_{55}+\rho_{22}\rho_{77}\\
&+\rho_{22}\rho_{88}+\rho_{44}\rho_{55}+\rho_{44}\rho_{77}+\rho_{44}\rho_{88}),
\end{aligned}
\end{eqnarray*}
\begin{eqnarray*}
\begin{aligned}
T^{2}_{b,-}&=\frac{9}{4}\rho_{44}\rho_{55}+\frac{3}{4}(\rho_{11}\rho_{55}+\rho_{22}\rho_{55}+\rho_{33}\rho_{55}\\
&+\rho_{44}\rho_{66}+\rho_{44}\rho_{77}+\rho_{44}\rho_{88})+\frac{1}{4}(\rho_{11}\rho_{66}\\
&+\rho_{11}\rho_{77}+\rho_{11}\rho_{88}+\rho_{22}\rho_{66}+\rho_{22}\rho_{77}\\
&+\rho_{22}\rho_{88}+\rho_{33}\rho_{66}+\rho_{33}\rho_{77}+\rho_{33}\rho_{88}),
\end{aligned}
\end{eqnarray*}
\begin{eqnarray*}
\begin{aligned}
T^{2}_{b,+}&=\frac{9}{4}\rho_{11}\rho_{88}+\frac{3}{4}(\rho_{11}\rho_{55}+\rho_{11}\rho_{66}+\rho_{11}\rho_{77}\\
&+\rho_{22}\rho_{88}+\rho_{33}\rho_{88}+\rho_{44}\rho_{88})+\frac{1}{4}(\rho_{22}\rho_{55}\\
&+\rho_{22}\rho_{66}+\rho_{22}\rho_{77}+\rho_{33}\rho_{55}+\rho_{33}\rho_{66}\\
&+\rho_{33}\rho_{77}+\rho_{44}\rho_{55}+\rho_{44}\rho_{66}+\rho_{44}\rho_{77}),
\end{aligned}
\end{eqnarray*}
\begin{eqnarray}
\begin{aligned}
\label{Ta-d2}
T^{2}_{a,+}&=\frac{9}{4}\rho_{22}\rho_{77}+\frac{3}{4}(\rho_{22}\rho_{55}+\rho_{22}\rho_{66}+\rho_{22}\rho_{88}\\
&+\rho_{11}\rho_{77}+\rho_{33}\rho_{77}+\rho_{44}\rho_{77})+\frac{1}{4}(\rho_{11}\rho_{55}\\
&+\rho_{11}\rho_{66}+\rho_{11}\rho_{88}+\rho_{33}\rho_{55}+\rho_{33}\rho_{66}\\
&+\rho_{33}\rho_{88}+\rho_{44}\rho_{55}+\rho_{44}\rho_{66}+\rho_{44}\rho_{88}).
\end{aligned}
\end{eqnarray}

Therefore, according to Eqs.(\ref{Ta})-(\ref{Ta-d2}), the steering from A to BC is given by
\begin{eqnarray*}
\begin{aligned}
\label{SBCA}
S^{A\rightarrow BC}=&max\left\{0, 4(|\rho_{18}|^{2}-T^{1}_{a,-}), 4(|\rho_{27}|^{2}-T^{1}_{b,-}),\right.\\
&\left.4(|\rho_{36}|^{2}-T^{1}_{b,+}), 4(|\rho_{45}|^{2}-T^{1}_{a,+})\right\},
\end{aligned}
\end{eqnarray*}
and the steering from BC to A is
\begin{eqnarray*}
\begin{aligned}
\label{S2BC-A}
S^{BC\rightarrow A}=&max\left\{0, \frac{16}{3}(|\rho_{18}|^{2}-T^{2}_{a,-}), \frac{16}{3}(|\rho_{27}|^{2}-T^{2}_{b,-}),\right.\\
&\left.\frac{16}{3}(|\rho_{36}|^{2}-T^{2}_{b,+}), \frac{16}{3}(|\rho_{45}|^{2}-T^{2}_{a,+})\right\}.
\end{aligned}
\end{eqnarray*}

Here, the coefficient $4$ and $\frac{16}{3}$ are set to guarantee that the steering of the maximally entangled state reaches $1$.

\section{APPENDIX B: The steering from B to CA and CA to B}
\label{App2}
Since $\rho_{ABC}=\rho_{X}$, we can get that
\begin{eqnarray}
\label{rhoCAB}
\rho_{CAB}=
\left (
\begin{array}{cccccccc}
\rho_{11}   &0               &0                  &0                 &0                &0                     &0                       &\rho_{18}\\
0           &\rho_{33}       &0                  &0                 &0                &0                     &\rho_{36}               &0        \\
0           &0               &\rho_{55}          &0                 &0                &\rho_{45}             &0                       &0        \\
0           &0               &0                  &\rho_{77}         &\rho_{27}        &0                     &0                       &0        \\
0           &0               &0                  &\rho_{72}         &\rho_{22}        &0                     &0                       &0        \\
0           &0               &\rho_{54}          &0                 &0                &\rho_{44}             &0                       &0        \\
0           &\rho_{63}       &0                  &0                 &0                &0                     &\rho_{66}               &0        \\
\rho_{81}   &0               &0                  &0                 &0                &0                     &0                       &\rho_{88}
\end{array}
\right ).
\end{eqnarray}

Thus, from Eq.(\ref{CAB}), we obtain that the steered density matrix $\tau^{1}_{CA|B}$ of $\rho_{CAB}$ can be clearly written as
\begin{widetext}
\begin{eqnarray}
\label{tauCAB}
\tau^{1}_{CA|B}=
\left (
\begin{array}{cccccccc}
\frac{\sqrt{3}}{3}\rho_{11}+i_{1}  &0               &0                  &0                 &0                &0                     &0            &\frac{\sqrt{3}}{3}\rho_{18}\\
0            &\frac{\sqrt{3}}{3}\rho_{33}+i_{1}     &0                  &0                 &0                &0                     &\frac{\sqrt{3}}{3}\rho_{36}           &0\\
0            &0               &\frac{\sqrt{3}}{3}\rho_{55}+i_{2}          &0                 &0                &\frac{\sqrt{3}}{3}\rho_{45}             &0                 &0\\
0            &0               &0            &\frac{\sqrt{3}}{3}\rho_{77}+i_{2}       &\frac{\sqrt{3}}{3}\rho_{27}          &0                     &0                       &0\\
0            &0               &0            &\frac{\sqrt{3}}{3}\rho_{72}         &\frac{\sqrt{3}}{3}\rho_{22}+i_{3}        &0                     &0                       &0\\
0            &0               &\frac{\sqrt{3}}{3}\rho_{54}    &0                 &0                &\frac{\sqrt{3}}{3}\rho_{44}+i_{3}             &0                       &0\\
0            &\frac{\sqrt{3}}{3}\rho_{63}       &0            &0                 &0                &0                  &\frac{\sqrt{3}}{3}\rho_{66}+i_{4}                  &0\\
\frac{\sqrt{3}}{3}\rho_{81}    &0               &0            &0                 &0                &0                  &0                     &\frac{\sqrt{3}}{3}\rho_{88}+i_{4}
\end{array}
\right ),
\end{eqnarray}
\end{widetext}
where $i_{1}=\frac{3-\sqrt{3}}{6}(\rho_{11}+\rho_{33})$, $i_{2}=\frac{3-\sqrt{3}}{6}(\rho_{55}+\rho_{77})$, $i_{3}=\frac{3-\sqrt{3}}{6}(\rho_{22}+\rho_{44})$, and $i_{4}=\frac{3-\sqrt{3}}{6}(\rho_{66}+\rho_{88})$.
The state $\tau^{1}_{BC|A}$ is entangled provided that it satisfies the inequality
\begin{eqnarray}
\begin{aligned}
\label{Qa}
|\rho_{18}|^{2}>Q^{1}_{a,+} \quad \quad \quad or \quad \quad \quad |\rho_{27}|^{2}>Q^{1}_{b,+},
\end{aligned}
\end{eqnarray}
or
\begin{eqnarray}
\begin{aligned}
\label{Qc}
|\rho_{36}|^{2}>Q^{1}_{a,-} \quad \quad \quad or \quad \quad \quad |\rho_{45}|^{2}>Q^{1}_{b,-},
\end{aligned}
\end{eqnarray}
where
\begin{eqnarray*}
\begin{aligned}
Q^{1}_{a,\pm}=&\frac{2\pm\sqrt{3}}{2}\rho_{22}\rho_{77}+\frac{2\mp\sqrt{3}}{2}\rho_{44}\rho_{55}\\
&+\frac{1}{2}(\rho_{22}\rho_{55}+\rho_{44}\rho_{77}),\\
\end{aligned}
\end{eqnarray*}
\begin{eqnarray}
\begin{aligned}
\label{Qa.d}
Q^{1}_{b,\pm}=&\frac{2\pm\sqrt{3}}{2}\rho_{11}\rho_{88}+\frac{2\mp\sqrt{3}}{2}\rho_{33}\rho_{66}\\
&+\frac{1}{2}(\rho_{11}\rho_{66}+\rho_{33}\rho_{88}).\\
\end{aligned}
\end{eqnarray}
Thus, the steering from B to CA is witnessed. Following that, we determine that the steered density matrix $\tau^{2}_{CA|B}$ associated with $\rho_{CAB}$ can be accurately represented as
\begin{widetext}
\begin{eqnarray}
\label{tau2CAB}
\tau^{2}_{CA|B}=
\left (
\begin{array}{cccccccc}
\frac{1}{3}\rho_{11}+j_{1}  &0               &0                  &0                 &0                &0                     &0            &\frac{1}{3}\rho_{18}\\
0            &\frac{1}{3}\rho_{33}+j_{2}     &0                  &0                 &0                &0                     &\frac{1}{3}\rho_{36}           &0\\
0            &0               &\frac{1}{3}\rho_{55}+j_{1}          &0                 &0                &\frac{1}{3}\rho_{45}             &0                 &0\\
0            &0               &0            &\frac{1}{3}\rho_{77}+j_{2}       &\frac{1}{3}\rho_{27}          &0                     &0                       &0\\
0            &0               &0            &\frac{1}{3}\rho_{72}         &\frac{1}{3}\rho_{22}+j_{1}        &0                     &0                       &0\\
0            &0               &\frac{1}{3}\rho_{54}    &0                 &0                &\frac{1}{3}\rho_{44}+j_{2}             &0                       &0\\
0            &\frac{1}{3}\rho_{63}       &0            &0                 &0                &0                  &\frac{1}{3}\rho_{66}+j_{1}                  &0\\
\frac{1}{3}\rho_{81}    &0               &0            &0                 &0                &0                  &0                     &\frac{1}{3}\rho_{88}+j_{2}
\end{array}
\right ),
\end{eqnarray}
\end{widetext}
where $j_{1}=\frac{1}{6}(\rho_{11}+\rho_{22}+\rho_{55}+\rho_{66})$ and $j_{2}=\frac{1}{6}(\rho_{33}+\rho_{44}+\rho_{77}+\rho_{88})$.
The state $\tau^{2}_{BC|A}$ is entangled whenever it satisfies the inequalities
\begin{eqnarray}
\begin{aligned}
\label{Qa2}
|\rho_{18}|^{2}>Q^{2}_{a,+} \quad \quad \quad or \quad \quad \quad |\rho_{27}|^{2}>Q^{2}_{b,+},
\end{aligned}
\end{eqnarray}
or
\begin{eqnarray}
\begin{aligned}
\label{Qc2}
|\rho_{36}|^{2}>Q^{2}_{a,-} \quad \quad \quad or \quad \quad \quad |\rho_{45}|^{2}>Q^{2}_{b,-},
\end{aligned}
\end{eqnarray}
where
\begin{eqnarray*}
\begin{aligned}
Q^{2}_{a,+}&=\frac{9}{4}\rho_{22}\rho_{77}+\frac{3}{4}(\rho_{11}\rho_{77}+\rho_{55}\rho_{77}+\rho_{66}\rho_{77}\\
&+(\rho_{22}\rho_{33}+\rho_{22}\rho_{44}+\rho_{22}\rho_{88})+\frac{1}{4}(\rho_{11}\rho_{33}\\
&+\rho_{11}\rho_{44}+\rho_{11}\rho_{88}+\rho_{33}\rho_{55}+\rho_{44}\rho_{55}\\
&+\rho_{55}\rho_{88}+\rho_{33}\rho_{66}+\rho_{44}\rho_{66}+\rho_{66}\rho_{88}),
\end{aligned}
\end{eqnarray*}
\begin{eqnarray*}
\begin{aligned}
Q^{2}_{b,+}&=\frac{9}{4}\rho_{11}\rho_{88}+\frac{3}{4}(\rho_{11}\rho_{33}+\rho_{11}\rho_{44}+\rho_{11}\rho_{77}\\
&+\rho_{22}\rho_{88}+\rho_{55}\rho_{88}+\rho_{66}\rho_{88})+\frac{1}{4}(\rho_{22}\rho_{33}\\
&+\rho_{22}\rho_{44}+\rho_{22}\rho_{77}+\rho_{33}\rho_{55}+\rho_{44}\rho_{55}\\
&+\rho_{55}\rho_{77}+\rho_{33}\rho_{66}+\rho_{44}\rho_{66}+\rho_{66}\rho_{77}),
\end{aligned}
\end{eqnarray*}
\begin{eqnarray*}
\begin{aligned}
Q^{2}_{a,-}&=\frac{9}{4}\rho_{44}\rho_{55}+\frac{3}{4}(\rho_{33}\rho_{55}+\rho_{55}\rho_{77}+\rho_{55}\rho_{88}\\
&+(\rho_{11}\rho_{44}+\rho_{22}\rho_{44}+\rho_{44}\rho_{66})+\frac{1}{4}(\rho_{11}\rho_{33}\\
&+\rho_{11}\rho_{77}+\rho_{11}\rho_{88}+\rho_{22}\rho_{33}+\rho_{22}\rho_{77}\\
&+\rho_{22}\rho_{88}+\rho_{33}\rho_{66}+\rho_{66}\rho_{77}+\rho_{66}\rho_{88}),
\end{aligned}
\end{eqnarray*}
\begin{eqnarray}
\begin{aligned}
\label{Qa.d2}
Q^{2}_{b,-}&=\frac{9}{4}\rho_{33}\rho_{66}+\frac{3}{4}(\rho_{11}\rho_{33}+\rho_{22}\rho_{33}+\rho_{33}\rho_{55}\\
&+\rho_{44}\rho_{66}+\rho_{66}\rho_{77}+\rho_{66}\rho_{88})+\frac{1}{4}(\rho_{11}\rho_{44}\\
&+\rho_{11}\rho_{77}+\rho_{11}\rho_{88}+\rho_{22}\rho_{44}+\rho_{22}\rho_{77}\\
&+\rho_{22}\rho_{88}+\rho_{44}\rho_{55}+\rho_{55}\rho_{77}+\rho_{55}\rho_{88}).
\end{aligned}
\end{eqnarray}

Furthermore, according to Eqs.(\ref{Qa})-(\ref{Qa.d2}), the steering from B to CA is given by
\begin{eqnarray*}
\begin{aligned}
\label{SCAB}
S^{B\rightarrow CA}=&max\left\{0, 4(|\rho_{18}|^{2}-Q^{1}_{a,+}), 4(|\rho_{27}|^{2}-Q^{1}_{b,+}),\right.\\
&\left.4(|\rho_{36}|^{2}-Q^{1}_{a,-}), 4(|\rho_{45}|^{2}-Q^{1}_{b,-})\right\},
\end{aligned}
\end{eqnarray*}
and the steering from CA to B is
\begin{eqnarray*}
\begin{aligned}
\label{S2CAB}
S^{CA\rightarrow B}=&max\left\{0, \frac{16}{3}(|\rho_{18}|^{2}-Q^{2}_{a,+}), \frac{16}{3}(|\rho_{27}|^{2}-Q^{2}_{b,+}),\right.\\
&\left.\frac{16}{3}(|\rho_{36}|^{2}-Q^{2}_{a,-}), \frac{16}{3}(|\rho_{45}|^{2}-Q^{2}_{b,-})\right\}.
\end{aligned}
\end{eqnarray*}

Here, the coefficient $4$ and $\frac{16}{3}$ are set to guarantee that the steering of the maximally entangled state reaches $1$.


\begin{thebibliography}{}

\bibitem{Einstein.1935}
A. Einstein, B. Podolsky, N. Rosen, Phys. Rev. \textbf{47}, 777 (1935).
\bibitem{Schrodinger.1935}
E. Schr\"{o}dinger, Math. Proc. Cambr. Philos. Soc. \textbf{31}, 555 (1935).
\bibitem{Schrodinger.1936}
E. Schr\"{o}dinger, Math. Proc. Cambr. Philos. Soc. \textbf{32}, 446 (1936).
\bibitem{Wiseman.2007}
H.M. Wiseman, S.J. Jones, A.C. Doherty, Phys. Rev. Lett. \textbf{98}, 140402 (2007).
\bibitem{Walborn.2011}
S.P. Walborn, A. Salles, R.M. Gomes, F. Toscano, P.H. SoutoRibeiro, Phys. Rev. Lett. \textbf{106}, 130402 (2011).
\bibitem{Navascues.2012}
M. Navascu\'{e}s, D. P\'{e}rez-Garc\'{\i}a, Phys. Rev. Lett. \textbf{109}, 160405 (2012).
\bibitem{Skrzypczyk.2014}
P. Skrzypczyk, M. Navascu\'{e}s, D. Cavalcanti, Phys. Rev. Lett. \textbf{112}, 180404 (2014).
\bibitem{Deng.2017}
X. Deng, Y. Xiang, C. Tian, G. Adesso, Q. He, Q. Gong, X. Su, C. Xie, K. Peng, Phys. Rev. Lett. \textbf{118}, 230501 (2017).
\bibitem{Nery.2018}
R.V. Nery, M.M. Taddei, R. Chaves, L. Aolita, Phys. Rev. Lett. \textbf{120}, 140408 (2018).
\bibitem{Nery.2020}
R.V. Nery, M.M. Taddei, P. Sahium, S.P. Walborn, L. Aolita, G.H. Aguilar, Phys. Rev. Lett. \textbf{124}, 120402 (2020).
\bibitem{Designolle.2021}
S. Designolle, V. Srivastav, R. Uola, N.H. Valencia, W. McCutcheon, M. Malik, N. Brunner, Phys. Rev. Lett. \textbf{126}, 200404 (2021).
\bibitem{Chen.2014}
Y.N. Chen, C.M. Li, N. Lambert, S.L. Chen, Y. Ota, G.Y. Chen, F. Nori, Phys. Rev. A \textbf{89}, 032112 (2014).
\bibitem{Kogias.2015}
I. Kogias, A.R. Lee, S. Ragy, G. Adesso, Phys. Rev. Lett. \textbf{114}, 060403 (2015).
\bibitem{Ren.2018}
C. Ren, H.Y. Su, H. Shi, J. Chen, Phys. Rev. A \textbf{97}, 032119 (2018).
\bibitem{Maity.2018}
A.G. Maity, S. Datta, A.S. Majumdar, Phys. Rev. A \textbf{96}, 052326 (2018).
\bibitem{Saunders.2010}
D.J. Saunders, S.J. Jones, H.M. Wiseman, G.J. Pryde, Nat. Phys. \textbf{6}, 845 (2010).
\bibitem{Wollmann.2016}
S. Wollmann, N. Walk, A.J. Bennet, H.M. Wiseman, G.J. Pryde, Phys. Rev. Lett. \textbf{116}, 160403 (2016).
\bibitem{Xiao.2017}
Y. Xiao, X.J. Ye, K. Sun, J.S. Xu, C.F. Li, G.C. Guo, Phys. Rev. Lett. \textbf{118}, 140404 (2017).

\bibitem{Branciard.2012}
C. Branciard, E.G. Cavalcanti, S.P. Walborn, V. Scarani, H.M. Wiseman, Phys. Rev. A \textbf{85}, 010301 (2012).
\bibitem{Handchen.2012}
V. H\"{a}ndchen, T. Eberle, S. Steinlechner, A. Samblowski, T. Franz, R.F. Werner, R. Schnabel, Nat. Photon. \textbf{6}, 596 (2012).
\bibitem{Armstrong.2015}
S. Armstrong, M. Wang, R.Y. Teh, Q. Gong, Q. He, J. Janousek, H.A. Bachor, M.D. Reid, P.K. Lam, Nat. Phys. \textbf{11}, 167 (2015).
\bibitem{Tischler.2018}
N. Tischler, F. Ghafari, T.J. Baker, S. Slussarenko, R.B. Patel, M.M. Weston, S. Wollmann, L.K. Shalm, V.B. Verma, S.W. Nam, H.C. Nguyen, H.M. Wiseman, G.J. Prydel, Phys. Rev. Lett. \textbf{121}, 100401 (2018).
\bibitem{Fan.2022}
Y. Fan, C. Jia, L. Qiu, Phys. Rev. A \textbf{106}, 012433 (2022).
\bibitem{Schwarzschild.1916}
K. Schwarzschild, Sitzungsberichte der k\"{o}niglich preussischen Akademie der Wissenschaften, 189 (1916).
\bibitem{Hawking.1974}
S.W. Hawking, Nature \textbf{248}, 30 (1974).
\bibitem{Hawking.1975}
S.W. Hawking, Commun. Math. Phys. \textbf{43}, 199 (1975).
\bibitem{Hawking.1976}
S.W. Hawking, Phys. Rev. D \textbf{14}, 2460 (1976).
\bibitem{Bombelli.1986}
L. Bombelli, R.K. Koul, J. Lee, R.D. Sorkin, Phys. Rev. D \textbf{34}, 373 (1986).
\bibitem{EHT.L1}
The Event Horizon Telescope Collaboration, Astrophys. J. Lett. \textbf{875}, L1 (2019).
\bibitem{EHT.L2}
The Event Horizon Telescope Collaboration, Astrophys. J. Lett. \textbf{875}, L2 (2019).
\bibitem{EHT.L3}
The Event Horizon Telescope Collaboration, Astrophys. J. Lett. \textbf{875}, L3 (2019).
\bibitem{EHT.L4}
The Event Horizon Telescope Collaboration, Astrophys. J. Lett. \textbf{875}, L4 (2019).
\bibitem{EHT.L5}
The Event Horizon Telescope Collaboration, Astrophys. J. Lett. \textbf{875}, L5 (2019).
\bibitem{EHT.L6}
The Event Horizon Telescope Collaboration, Astrophys. J. Lett. \textbf{875}, L6 (2019).
\bibitem{EHT.L17}
The Event Horizon Telescope Collaboration, Astrophys. J. Lett. \textbf{930}, L17 (2022).
\bibitem{Schuller.2005}
I. Fuentes-Schuller, R.B. Mann, Phys. Rev. Lett. \textbf{95}, 120404 (2005).
\bibitem{Pan.2008}
Q. Pan, J. Jing, Phys. Rev. D \textbf{78}, 065015 (2008).
\bibitem{Wang.2009}
J. Wang, Q. Pan, S. Chen, J. Jing, Phys. Lett. B \textbf{677}, 186 (2009).
\bibitem{Wang.2010}
J. Wang, Q. Pan, J. Jing, Phys. Lett. B \textbf{692}, 202 (2010)
\bibitem{Esfahani.2011}
B. Nasr Esfahani, M. Shamirzaie, M. Soltani, Phys. Rev. D \textbf{84}, 025024 (2011).
\bibitem{Bhattacharya.2022}
S. Bhattacharya, N. Joshi, Phys. Rev. D \textbf{105}, 065007 (2022)
\bibitem{Wu.2022}
S.M. Wu, H.S. Zeng, Eur. Phys. J. C \textbf{82}, 4 (2022).
\bibitem{Li.2022}
L.J. Li, F. Ming, X.K. Song, L. Ye, D. Wang, Eur. Phys. J. C \textbf{82}, 726 (2022).
\bibitem{Zhang.2023}
T. Zhang, X. Wang, S.M. Fei, Eur. Phys. J. C \textbf{83}, 607 (2023).
\bibitem{Mi.2025}
G.W. Mi, X. Huang, S.M. Fei, T. Zhang, Eur. Phys. J. C \textbf{85}, 354 (2025).
\bibitem{Dong.2018}
W.C. Qiang, G.H. Sun, Q. Dong, S.H. Dong, Phys. Rev. A \textbf{98}, 022320 (2018).
\bibitem{Dong.2019}
A.J. Torres-Arenas, Q. Dong, G.H. Sun, W.C. Qiang, S.H. Dong, Phys. Lett. B \textbf{789}, 93 (2019). 
\bibitem{Dong.2025}
A.R.P. Moreira, A. Bouzenada, S.H. Dong, G.H. Sun, F. Ahmed, Eur. Phys. J. C \textbf{85}, 1067 (2025).
\bibitem{D.Das.2019}
D. Das, S. Sasmal, S. Roy, Phys. Rev. A \textbf{99}, 052109 (2019).
\bibitem{EPJC.2022}
S.M. Wu, H.S. Zeng, Eur. Phys. J. C \textbf{82}, 716 (2022).
\bibitem{Fei.2022}
Z. Chen, S.M. Fei, Entropy \textbf{24}, 1297 (2022).
\bibitem{Wu.2025}
S.H. Shang , S.M. Wu, Eur. Phys. J. C \textbf{85}, 790 (2025).
\bibitem{Dirac.1957}
D.R. Brill, J.A. Wheeler, Rev. Mod. Phys. \textbf{29}, 465 (1957).
\bibitem{metric.2010}
E. Mart\'in-Mart\'inez, L. J. Garay, and J. Le\'on, Phys. Rev. D \textbf{82}, 064006 (2010).
\bibitem{Xu.2014}
S. Xu, X. K. Song, J. D. Shi, and L. Ye, Phys. Rev. D \textbf{89}, 065022 (2014).
\bibitem{Jing.2004}
J.L. Jing, Phys. Rev. D \textbf{70} 065004 (2004).
\bibitem{Damoar.1976}
T. Damoar, R. Ruffini, Phys. Rev. D \textbf{14}, 332 (1976)
\bibitem{Barnett.1997}
S.M. Barnett, P.M. Radmore, Oxford University Press, New York (1997).
\bibitem{Wang2010}
J. Wang, Q. Pan, J. Jing, Ann. Phys. \textbf{325}, 1190 (2010).
\bibitem{Phy2024}
S.M. Wu, X.W. Teng, J.X. Li, S.H. Li, T.H. Liu, and J.C. Wang, Phy. Lett. B \textbf{848}, 138334 (2024).
\bibitem{PRD.2016}
J. Wang, H. Cao, J. Jing, H. Fan, Phys. Rev. D \textbf{93}, 125011 (2016).
\bibitem{APB.2018}
J. Wang, J. Jing, H. Fan, Ann. Phys. Berlin \textbf{530}, 1700261 (2018).


\end{thebibliography}
\end{document}